\documentclass[twocolumn,showpacs,preprintnumbers,amssymb,amsmath,superscriptaddress,letterpaper]{revtex4}[10pt]

\usepackage{graphicx}
\usepackage{dcolumn}
\usepackage{bm}
\usepackage{pstricks}
\usepackage{color}
\usepackage{epsfig}

\newcommand{\be}{\begin{equation}}
\newcommand{\ee}{\end{equation}}
\newcommand{\bea}{\begin{eqnarray}}
\newcommand{\eea}{\end{eqnarray}}

\newcommand{\gsim}{ \mathop{}_{\textstyle \sim}^{\textstyle >} }

\newcommand{\cm}{{\rm ~cm}}
\newcommand{\cmcubps}{{\rm ~cm^3/s}}

\newcommand{\MeV}{{\rm ~MeV}}
\newcommand{\GeV}{{\rm ~GeV}}
\newcommand{\gev}{{\rm ~GeV}}

\newcommand{\sigmav}{\langle\sigma_Av\rangle}

\newcommand{\epp}{e^\pm}

\begin{document}

\title{The Fermi gamma-ray spectrum of the inner galaxy:\\
 Implications for annihilating dark matter}

\author{Ilias Cholis}
\affiliation{Center for Cosmology and Particle Physics, Department of Physics, New York University, 
New York, NY 10003}

\author{Gregory Dobler}
\affiliation{Harvard-Smithsonian Center for Astrophysics, 60 Garden St., Cambridge, MA 02138}

\author{Douglas P. Finkbeiner}
\affiliation{Harvard-Smithsonian Center for Astrophysics, 60 Garden St.,
  Cambridge, MA 02138}
\affiliation{Physics Department, Harvard University, Cambridge, MA 02138}

\author{Lisa Goodenough}
\affiliation{Center for Cosmology and Particle Physics, Department of Physics, New York University, 
New York, NY 10003}

\author{Tracy R. Slatyer}
\affiliation{Physics Department, Harvard University, Cambridge, MA 02138}

\author{Neal Weiner}
\affiliation{Center for Cosmology and Particle Physics, Department of Physics, New York University, 
New York, NY 10003}

\date{\today}

\begin{abstract}

  Recently, a preliminary spectrum from the Fermi Gamma-ray 
  Space Telescope has been presented for the inner galaxy 
  ($- 30^\circ < \ell < 30^\circ$, $-5^\circ < b < 5^\circ$), 
  as well as the galactic center ($-1^\circ < \ell < 1^\circ$, $- 
  1^\circ < b < 1^\circ$).  We consider the implications of 
  these data for dark matter annihilation models, especially 
  models capable of producing the cosmic-ray excesses 
  previously observed by PAMELA and Fermi.  These local 
  cosmic-ray excesses, when extrapolated to the inner galaxy, imply
  inverse Compton scattering (ICS) gamma-ray signals largely
  consistent with the preliminary Fermi gamma-ray spectrum. For
  specific halos and models, particularly those with prompt photons,
  the data have begun to constrain the allowed parameter space.
  Although significant modeling and background uncertainties remain,
  we explore how large a signal is permitted by the current
  data. Based upon this, we make
  predictions for what signal could be present in other 
  regions of the sky where dark matter signals may be easier 
  to isolate from the astrophysical backgrounds.
\end{abstract}
\maketitle

\section{Introduction}
\label{sec:intro}
During its first year of operation, the Fermi Gamma-ray Space Telescope
\cite{Gehrels:1999ri}\footnote{The Fermi homepage is
  \texttt{http://fermi.gsfc.nasa.gov}} has revolutionized the study of
gamma-ray pulsars \cite{Abdo:2008pulsar} and transients
\cite{Abdo:2009transients}, and has determined the $\epp$ cosmic-ray
spectrum up to 1 TeV \cite{Abdo:2009zk} with high statistics.  The
first-year data release (expected 11 Aug 2009) promises further striking
advances: the high angular and energy resolution of the Large Area
Telescope (LAT) will allow Fermi to explore the $\sim$ 100 GeV sky,
providing a full-sky map of the ISM $\pi^0$ and inverse Compton
components for the first time.  Among the many objectives of the mission
is the search for gamma rays produced by dark matter annihilation or
decay\footnote{See p. 15 of the Fermi Science Requirements Document at
  \texttt{http://fermi.gsfc.nasa.gov/science/\\433-SRD-0001\_CH-04.pdf}}.

Gamma rays have long been recognized as a natural consequence of
Weakly Interacting Massive Particle (WIMP) models (e.g.,
\cite{Srednicki:1985sf,Bergstrom:1988fp,Rudaz:1987ry}), and Fermi was
expected to probe these models \cite{glast} by studying galactic
subhalos, the inner galaxy, and the extragalactic background.
Conventional WIMPs, such as the neutralino, produce copious gamma rays
from $\pi^0$ decays in the hadronic cascade \cite{Jungman:1995df}.

The situation has changed significantly in the past year, however. The PAMELA satellite \cite{Adriani:2008zr} has found evidence for a new primary source of 10-100 GeV positrons, while the Fermi \cite{Abdo:2009zk}, ATIC \cite{aticlatest} and H.E.S.S experiments \cite{Aharonian:2009ah} have detected a hardening in the local $e^+ + e^-$ spectrum at $\sim 300-1000$ GeV \footnote{Alternative explanations without new primary sources have also been proposed, such as secondary production in supernova remnants \cite{Blasi:2009hv} and modifications to electron propagation \cite{Katz:2009yd}.}. While e.g. pulsars provide an astrophysical candidate for this new source \cite{pulsars,pulsars2,2001A&A...368.1063Z,Hooper:2008kg,Yuksel:2008rf,Profumo:2008ms,Malyshev:2009tw,Kawanaka:2009dk}, dark matter is also an exciting possibility.

Thermal relic dark matter, which through annihilations was in equilibrium with the photon bath in the early universe, is expected to produce cosmic ray antimatter in the galactic halo through the same process. However, the observed $e^+ e^-$ excess is difficult to incorporate into most conventional WIMP models because they generally 1) produce too few positrons, 2) produce too soft a spectrum and 3) produce too many antiprotons, which have shown no signs of an excess \cite{Adriani:2008zq}. While one can by fiat insist that annihilations are dominantly into leptons, the large cross section is unexplained.

This problem can be solved by invoking a new force in the dark sector
so that dark matter interacts via -- and annihilates into -- a new
$\sim$ GeV scale force carrier \cite{ArkaniHamed:2008qn}. Such models,
(which we refer to as ``exciting dark matter'' or XDM), have been previously proposed in
the context of the 511 keV excess measured by
INTEGRAL \cite{Finkbeiner:2007kk}\footnote{Such models have also been grouped into the broader category of what has been termed ``secluded dark matter'' \cite{Pospelov:2007mp}, which includes annihilations into standard model singlet bosons or fermions of mass scale $m < m_\chi$.}. Because the annihilations are into light states, hard lepton spectra are naturally produced, but antiproton production is kinematically forbidden \cite{Cholis:2008vb,Cholis:2008qq}. The light force carrier yields an enhancement to the annihilation cross section through the Sommerfeld enhancement \cite{ArkaniHamed:2008qn}\footnote{The SE was originally studied in \cite{sommerfeld} and first discussed in the context of DM in \cite{Hisano:2004ds}.} or through capture into WIMPonium \cite{Pospelov:2008jd,MarchRussell:2008tu}.
(For various other models, see \cite{Chen:2008yi,Nelson:2008hj,Cholis:2008qq,Nomura:2008ru,Harnik:2008uu,Bai:2008jt,Fox:2008kb,Ponton:2008zv,Chen:2008qs,Ibe:2008ye,Chun:2008by,Arvanitaki:2008hq,Grajek:2008pg,Shirai:2009kh,Mardon:2009gw}.) At the same time, $\pi^0$'s are generally not copiously produced in these
annihilations, and so prompt photons are limited to those from final
state radiation (FSR) \cite{Beacom:2004pe,Bergstrom:2004cy,Birkedal:2005ep,Mack:2008wu,Bergstrom:2008gr,Bertone:2008xr,Bergstrom:2008ag,Meade:2009rb,Mardon:2009rc,Meade:2009iu}, which are suppressed compared to electronic production. 

However, shortly after the announcement of the electronic excesses, it was argued \cite{Cholis:2008wq} that a diffuse gamma ray signal from inverse Compton scattering (on starlight, far-infrared emission from dust, and the CMB) would be produced from essentially any model of dark matter annihilation that explains the electronic excesses, and provide ``smoking gun'' evidence of its origin (see also \cite{Zhang:2008tb,Borriello:2009fa,Cirelli:2009vg,Regis:2009md,Belikov:2009cx,Meade:2009iu} for discussions of the ICS signal). The dominant process by which electrons lose energy in the inner galaxy is ICS, so the energy of annihilation is largely converted into high energy gamma rays. However, the electrons can propagate away from the point of production before scattering, so these photons are diffused: at the energies in question, electrons and positrons propagate $\sim$ 1 kpc, and consequently the signal is expected to be spread over $\sim 5^\circ - 10^\circ$, irrespective of the cuspiness of the profile. This is in sharp contrast with FSR signals or other photons produced promptly in DM annihilations (such as $\pi^0$ gammas).

Recently, the Fermi-LAT collaboration has presented preliminary
measurements of gamma rays from the inner galaxy (IG) ($|\ell| < 30^\circ$, $ |b| < 5^\circ$) \cite{TROYTALK,DRELLTALK} and galactic center (GC)
($|\ell| < 1^\circ$, $|b| < 1^\circ$) \cite{SIMONATALK}\footnote{See
  \texttt{http://www-conf.slac.stanford.edu/tevpa09}}.  Although the GC
is background dominated, there is room in the data for a DM signal.  In
the inner galaxy, in particular, there seems room for some additional contribution above $\gsim 20 \gev$, when compared to expected backgrounds.  In this paper, we have
assumed that the background from instrumental and extragalactic backgrounds presented in \cite{TROYTALK} up to 75 GeV is
flat in $E^2 dN/dE$ up to higher energies.  This extrapolated background
is an order of magnitude below the data points, and preliminary
estimates \cite{ACKERMANNTALK} are that it could be even lower. 

There is a long list of caveats that must be applied to any 
analysis of these data.  First, because this angular range 
is optimized for emission mechanisms with a disk-like 
morpholgy, it is precisely the {\em wrong} region to attempt 
to claim a definitive signal of dark matter which should be 
more spherical \footnote{The situation is a bit more complex 
since the geometry of the DM ICS signal will depend on the 
specific \emph{shape} of the DM halo: ellipticity, 
concentration, etc.  Furthermore, the ICS geometry is also 
dependent on the ISRF geometry which is more disk-like.  We 
explore halo morphology in \cite{dobler:2009zz}}. In 
addition, since the DM density is significantly higher in 
the inner $10^\circ$, the S/B for a DM signal is expected to 
be significantly higher with a smaller angular window.  
Moreover, it must be emphasized that the preliminary 
spectrum may shift, estimates of the instrumental and 
isotropic backgrounds will be refined, and expectations for 
gamma production in the inner galaxy (especially from point 
sources) will be reexamined. In particular, there are known 
contaminants at high energies from charged cosmic rays 
\cite{simonaprivate} and the LAT team is working to remove 
them.

Nonetheless, in models that explain the Fermi/PAMELA excesses, the expected electronic production is so large that an appreciable signal could arise in this region. We thus focus here on the dark matter gamma ray signal in this angular range. We will show that the data are consistent with expectations from dark matter models for Fermi/PAMELA, and for some halo models, a significant fraction of the signal at high energies could arise from DM. 

In the following sections we investigate how large a dark matter
annihilation signal could be present, without conflicting with the preliminary data. Proceeding from this normalization, we make predictions for other regions of the sky. We will see that large signals above background are expected in the inner $5^\circ$ as well as in the ``four corners'' region ($5^\circ < |\ell|< 10^\circ $, $5^\circ < |b|< 10^\circ $). Because some
of the signal results from point sources in the Galactic plane,
these estimates of the dark matter contribution should be taken as
  upper limits.  Only careful modeling with the full data set will be
able to establish that the signal considered here is concentrated in the
inner galaxy, as expected, and is not found throughout the galactic
plane. 

\section{Photons from Dark Matter Annihilations}
Photons from dark matter can arise ``promptly'' or through inverse Compton scattering (ICS) processes. Prompt photons arise when photons are actually a component of the annihilation process, for example $\chi \chi \rightarrow \gamma \gamma$ or $\chi \chi \rightarrow {\rm hadrons}$, where $\pi^0$'s in the final state decay to photons. Such a signal traces $\rho^2$ \footnote{There is an exception in cases when the annihilation products are extremely long-lived \cite{Rothstein:2009pm}.}, i.e., it traces the dark matter annihilation itself.

In contrast, if the DM annihilation contains many hard electrons and positrons, then there is a signal of diffuse emission, from the interactions of the $\epp$ with the ISM. Indeed, such a signal is expected as a ``smoking gun'' of models to explain Fermi and PAMELA \cite{Cholis:2008wq}.

For a highly relativistic electron scattering on low energy photons, the spectrum of upscattered photons is given by \cite{1970RvMP...42..237B},
\bea \nonumber \frac{dN}{dE_\gamma d\epsilon dt} &=& \frac{3}{4} \sigma_T c \frac{(m_e c^2)^2}{\epsilon E_e^2} \big(2 q \log q + (1 + 2 q) (1 - q) \\ &&+ 0.5 (1 - q) (\Gamma q)^2 / (1 + \Gamma q) \big) n(\epsilon) \label{eq:klein-nishina}, 
\eea
\[ \Gamma =  4 \epsilon E_e / (m_e c^2)^2, \quad q = \frac{E_\gamma}{E_e} \frac{1}{\Gamma (1 - E_\gamma/E_e)},\]
\[\epsilon < E_\gamma < E_e \Gamma / (1 + \Gamma). \]
Here $\epsilon$ is the initial photon energy, $E_e$ is the electron energy, $E_\gamma$ is the energy of the upscattered gamma ray, and $n(\epsilon)$ describes the energy distribution of the soft photons per unit volume. Where $\Gamma \ll 1$, in the Thomson limit, the average energy of the upscattered photons is given by,
\begin{equation} \langle E_\gamma \rangle = (4/3) \gamma^2 \langle \epsilon \rangle, \end{equation}
where $\gamma = E_e / m_e c^2$ is the Lorentz boost of the electron. In the Klein-Nishina limit, $\Gamma \gg 1$, the spectrum instead peaks at the high energy end, and the upscattered photon carries away almost all the energy of the electron. For inverse Compton scattering on starlight ($\epsilon \sim {\rm 1 eV}$), $\Gamma \sim 1$ corresponds to $E_e \sim 65$ GeV: consequently, if the 300+ GeV photons observed by Fermi originate by inverse Compton scattering on starlight, they indicate the presence of electrons of similar energy.

Remarkably, we already have evidence for excess electronic production in
high energy electrons from two other sources: local cosmic rays and
excess microwaves in the inner galaxy. These measurements are
complementary with Fermi LAT measurements of the ICS spectrum: combining
them might allow us to test the synchrotron hypothesis for the origin of
the haze and potentially the hypothesis that the cosmic ray excesses
are due to local sources.

The ATIC balloon experiment \cite{aticlatest} has measured the spectrum of $e^++e^-$ (ATIC
cannot distinguish positrons from electrons) from 20-2000 GeV, and finds a broad excess at $300-800$ GeV, in agreement with the similar excess
observed by PPB-BETS
\cite{Torii:2008xu}. The Fermi \cite{Abdo:2009zk} and H.E.S.S \cite{Aharonian:2009ah} experiments have measured a similar but somewhat smaller $e^+ + e^-$ excess in the $300-1000$ GeV energy range, relative to the standard diffusive propagation model, and have not confirmed the peak and sharp cutoff observed by ATIC around 700 GeV. The Fermi electron spectrum can be well described as a power law with a spectral index $\sim -3.04$ in the $20-1000$ GeV range, which is significantly harder than predicted by conventional models for the electron spectrum.

The Wilkinson Microwave Anisotropy Probe (WMAP 
\cite{Bennett:2003ca}) has produced full sky maps from 23-94 
GHz \cite{Hinshaw:2003ex} which have provided both strong 
constraints on the cosmological parameters 
\cite{Spergel:2006hy} from the cosmic microwave background 
(CMB) anisotropies as well as exquisite measurements of 
Galactic microwave emission mechanisms at degree angular 
scales.  Recently, \cite{Finkbeiner:2003im,Dobler:2007wv} 
showed that, in addition to the four commonly accepted 
diffuse emission mechanisms (thermal bremsstrahlung from hot 
gas, thermal dust emission from interstellar dust grains, 
spinning dust emission from the smallest grains, and 
synchrotron from supernova shock accelerated electrons), 
there is a fifth Galactic component of diffuse synchrotron 
centered roughly on the Galactic center (GC) and extending 
for $\sim25$ degrees.  This emission has been termed the 
WMAP ``haze'' \cite{Finkbeiner:2003im,Dobler:2007wv}.  It 
has been pointed out that the geometry and spectrum of the 
haze are consistent with synchrotron from injection of 
$e^+e^-$ products of dark matter annihilation within the 
Galactic halo \cite{Finkbeiner:2004us,Hooper:2007kb} and 
that these same $e^+e^-$ should give rise to an ICS signal 
towards the GC in the Fermi data 
\cite{Cholis:2008wq}.  In this paper we concentrate on the 
particle physics models which give rise to an ICS signal 
that is consistent with the preliminary Fermi gamma-ray 
spectrum; a study of the consistency of the Fermi spectrum 
with the WMAP haze (including astrophysical and haze 
modeling uncertainties) in the context of DM annihilation 
will be presented in \cite{dobler:2009zz}.



\begin{figure}
\begin{center}
\includegraphics[width=.45\textwidth]{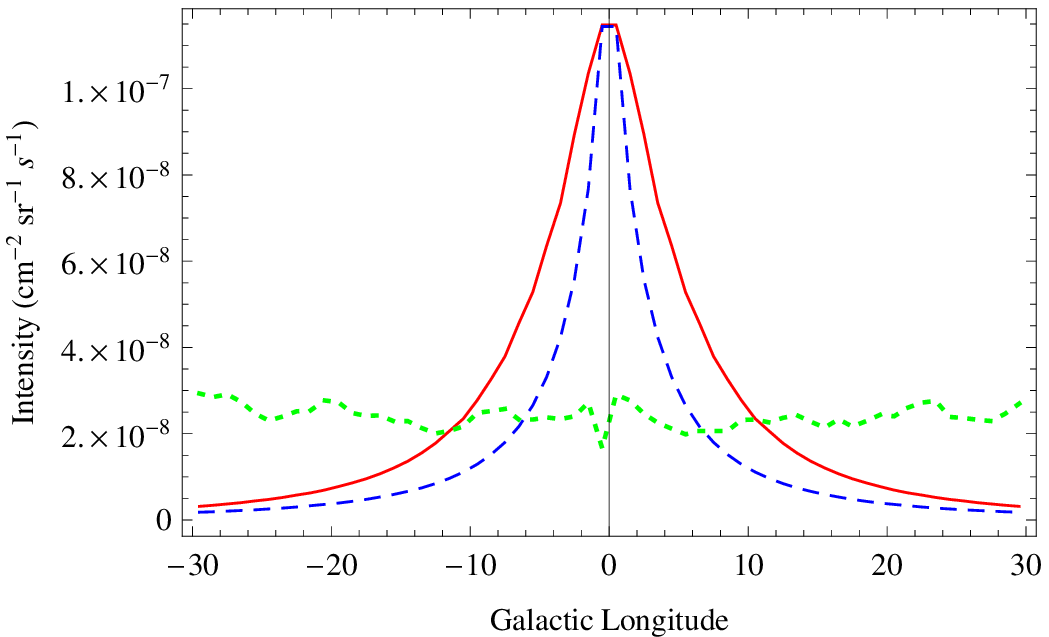}
\includegraphics[width=.45\textwidth]{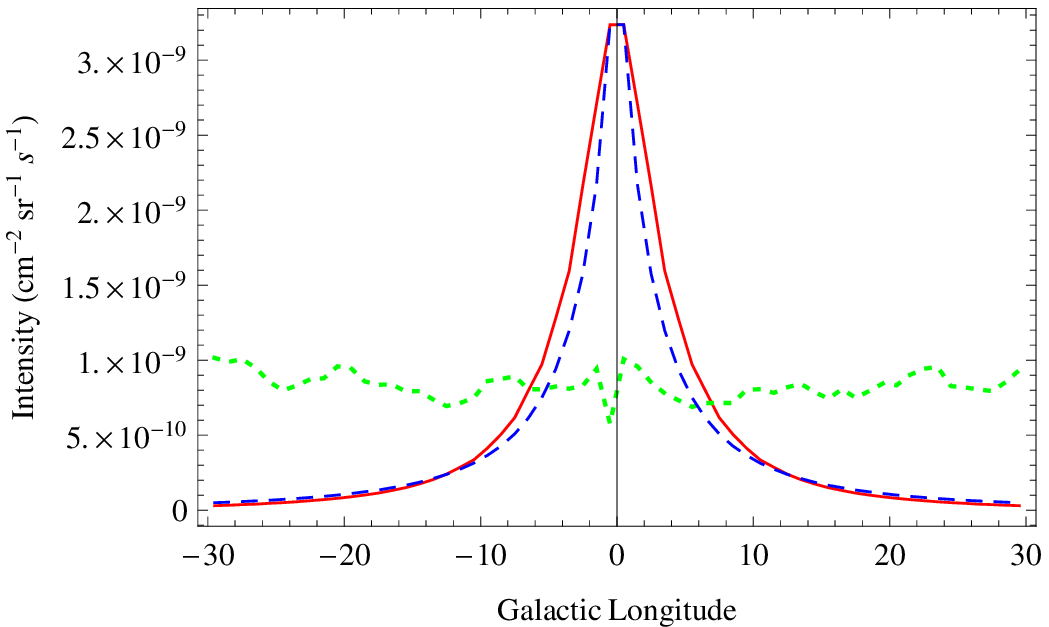}
\end{center}
\caption{The contribution to the total inner galaxy signal as a function of longitude for prompt photons (blue, long-dashed) and ICS photons (red, solid) and $\pi^0$ background (green, short-dashed). Plotted is the ICS signal from XDM muons with a mass of 3 TeV and $B_{0.4}=340$, and a prompt photon signal normalized to the ICS signal at $\ell = 0$. The sky distribution for 100 GeV photons (top) is broader for ICS than for 750 GeV photons (bottom).}
\label{fig:skydist}
\end{figure}

ICS and synchrotron signals both arise from electrons which may have diffused away from their production point. Consequently, as we have noted above, the sky distribution of the prompt and ICS signals is different. We show in Figure \ref{fig:skydist} the fraction of the total inner galaxy signal coming in $|b|< 5^\circ$ as a function of $\ell$. While the distribution of the diffuse ICS signal depends on propagation parameters, it is still generally very broad in comparison (see \cite{Borriello:2009fa} for a discussion of the sky distribution of ICS signals) . Both components, if present, can be important when viewing the inner galaxy.

\section{Dark Matter Models and ICS in the Inner Galaxy}
We calculate the spectrum of gamma rays using the GALPROP package \cite{galprop,Strong:1999sv,Galprop1}. For details, see \cite{Cholis:2008wq}. We normalize the dark matter distribution to a reference local dark matter density $\rho_0 = 0.4 \GeV \cm^{-3}$, following \cite{Catena:2009mf}, so our boosts are different by a factor of $0.56$ from those appearing in papers that consider $\rho_0 = 0.3 \GeV \cm^{-3}$. We consider six principle annihilation channels, i) direct to muons ($\chi \chi \rightarrow \mu^+ \mu^-$), ii) XDM electrons  ($\chi \chi \rightarrow \phi \phi$, $\phi \rightarrow \mu^+ \mu^-$) , iii) XDM muons ($\chi \chi \rightarrow \phi \phi$, $\phi \rightarrow \mu^+ \mu^-$) , iv) XDM $e^+e^-\mu^+\mu^-\pi^+\pi^-$ 1:1:2 ($\chi \chi \rightarrow \phi \phi$, $\phi \rightarrow \epp, \, \mu^+ \mu^-,\, \pi^+\pi^-$ in a 1:1:2 ratio), v) XDM taus  ($\chi \chi \rightarrow \phi \phi$, $\phi \rightarrow \tau^+ \tau^-$), and vi) direct annihilation to W's ($\chi \chi \rightarrow W^+ W^-$). The first four generate the bulk of their gammas from ICS processes, with a small FSR component. The final two have a significant prompt photon component as well. Model (iv) can arise where the $\phi$ is a gauge boson and kinetically mixes with the photon and $\rho$ meson at intermediate ($\sim 650 \MeV$) masses. See \cite{Meade:2009rb} for a discussion. Note that the signal of (iii) is essentially that of \cite{Nomura:2008ru} without the additional decays from the scalar state, but is qualitatively very similar.

\begin{figure*}
\begin{center}
\includegraphics[width=.45\textwidth]{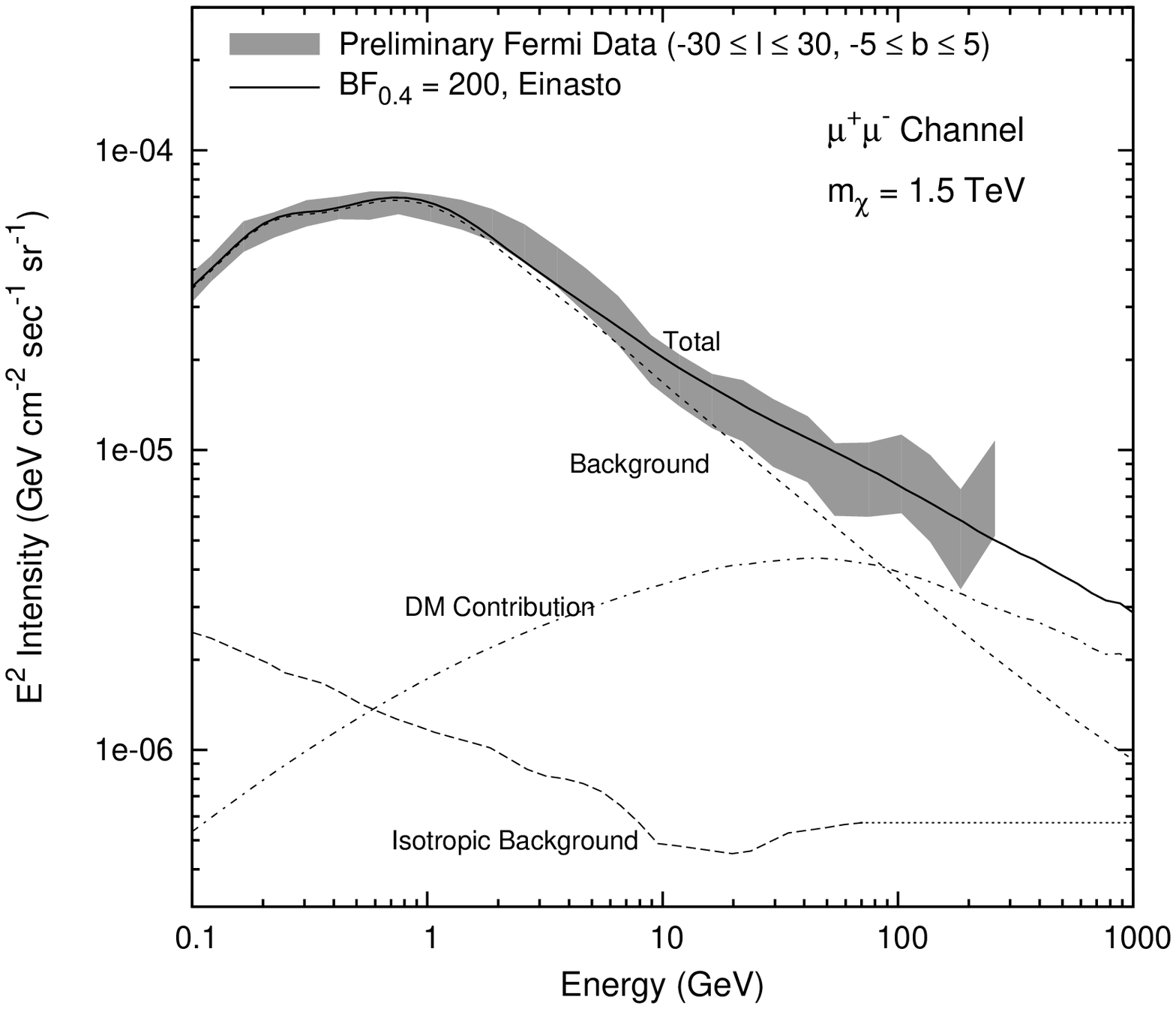}\hskip 0.2in
\includegraphics[width=.45\textwidth]{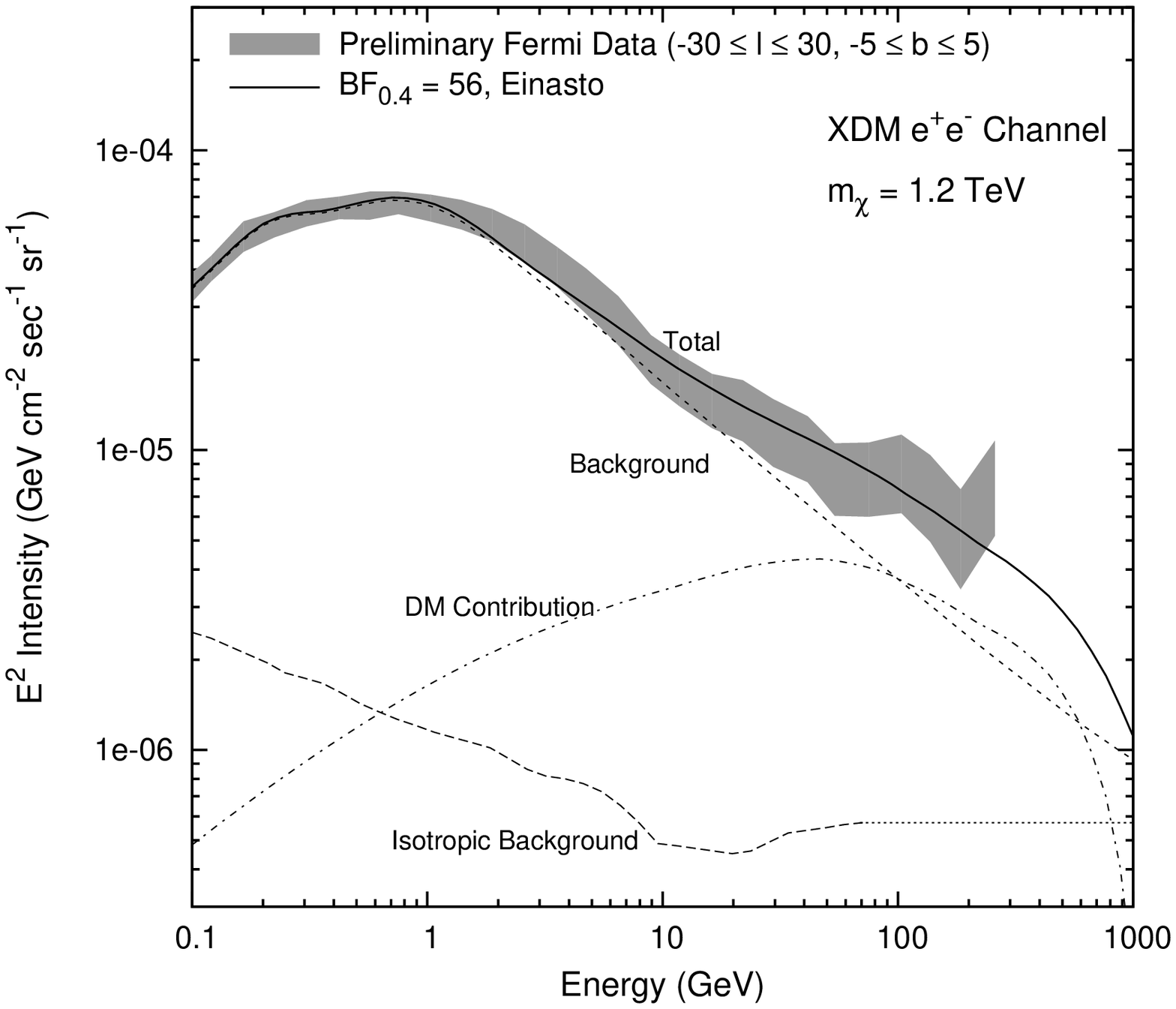}\\
\includegraphics[width=.45\textwidth]{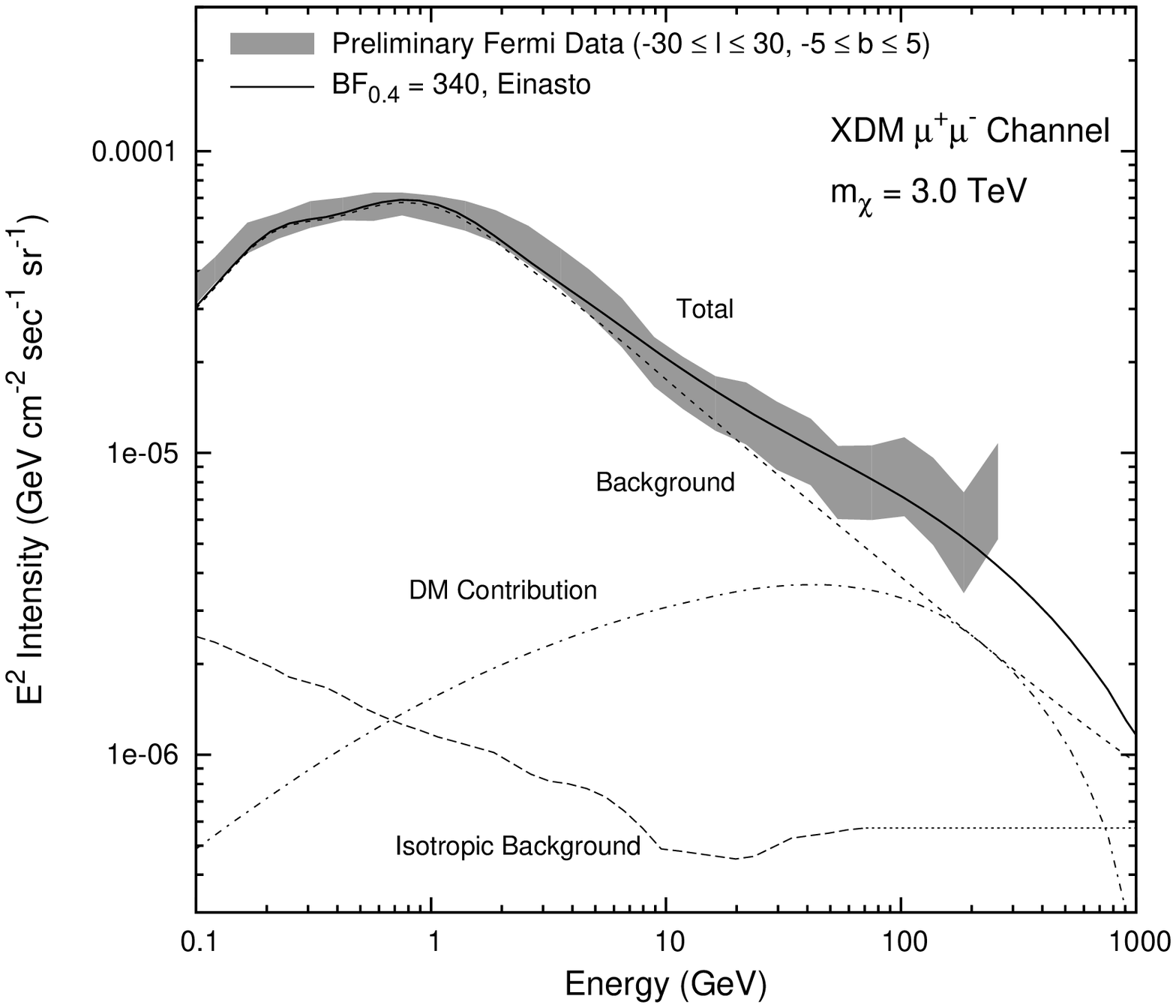}\hskip 0.2in
\includegraphics[width=.45\textwidth]{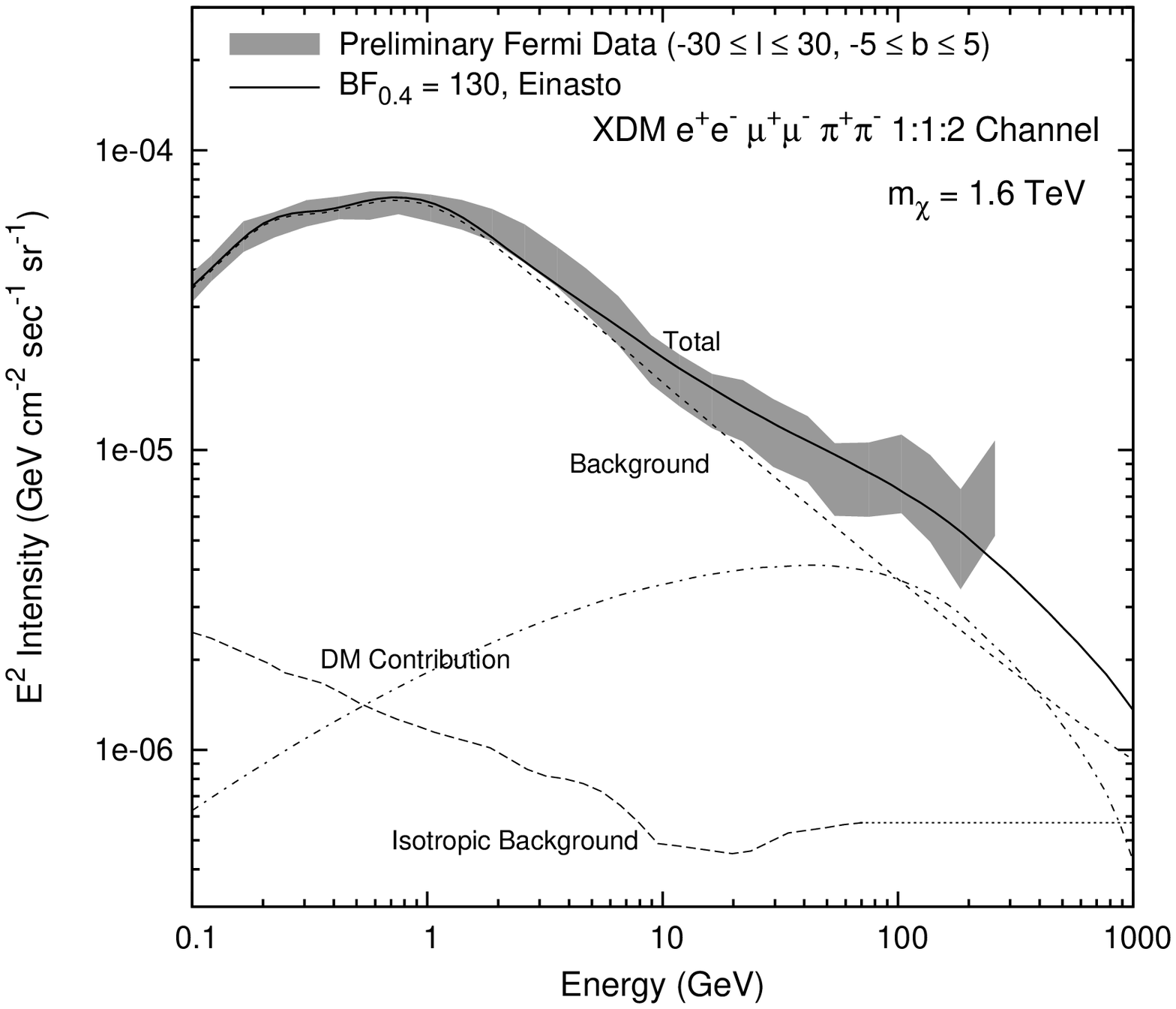}
\includegraphics[width=.45\textwidth]{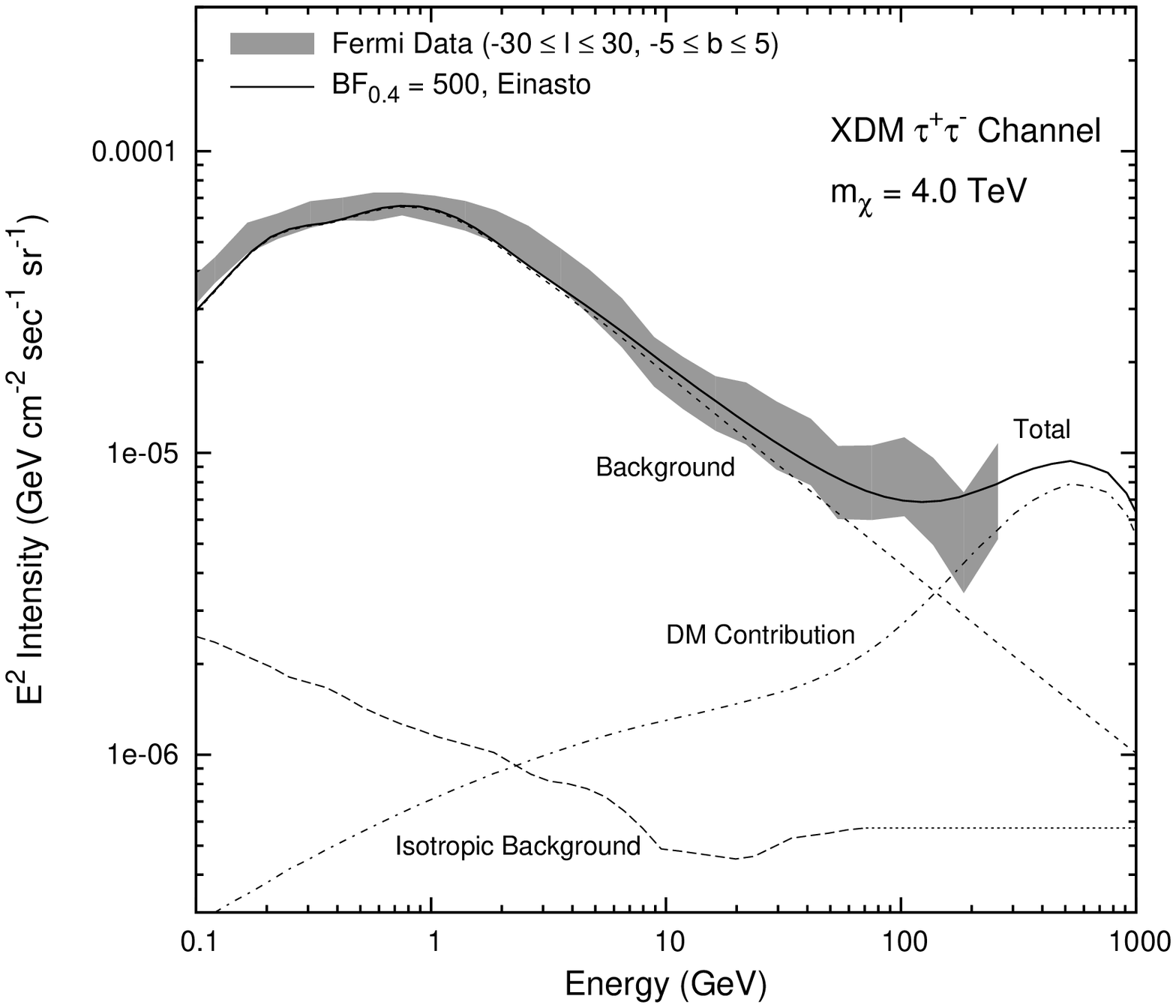}\hskip 0.2in
\includegraphics[width=.45\textwidth]{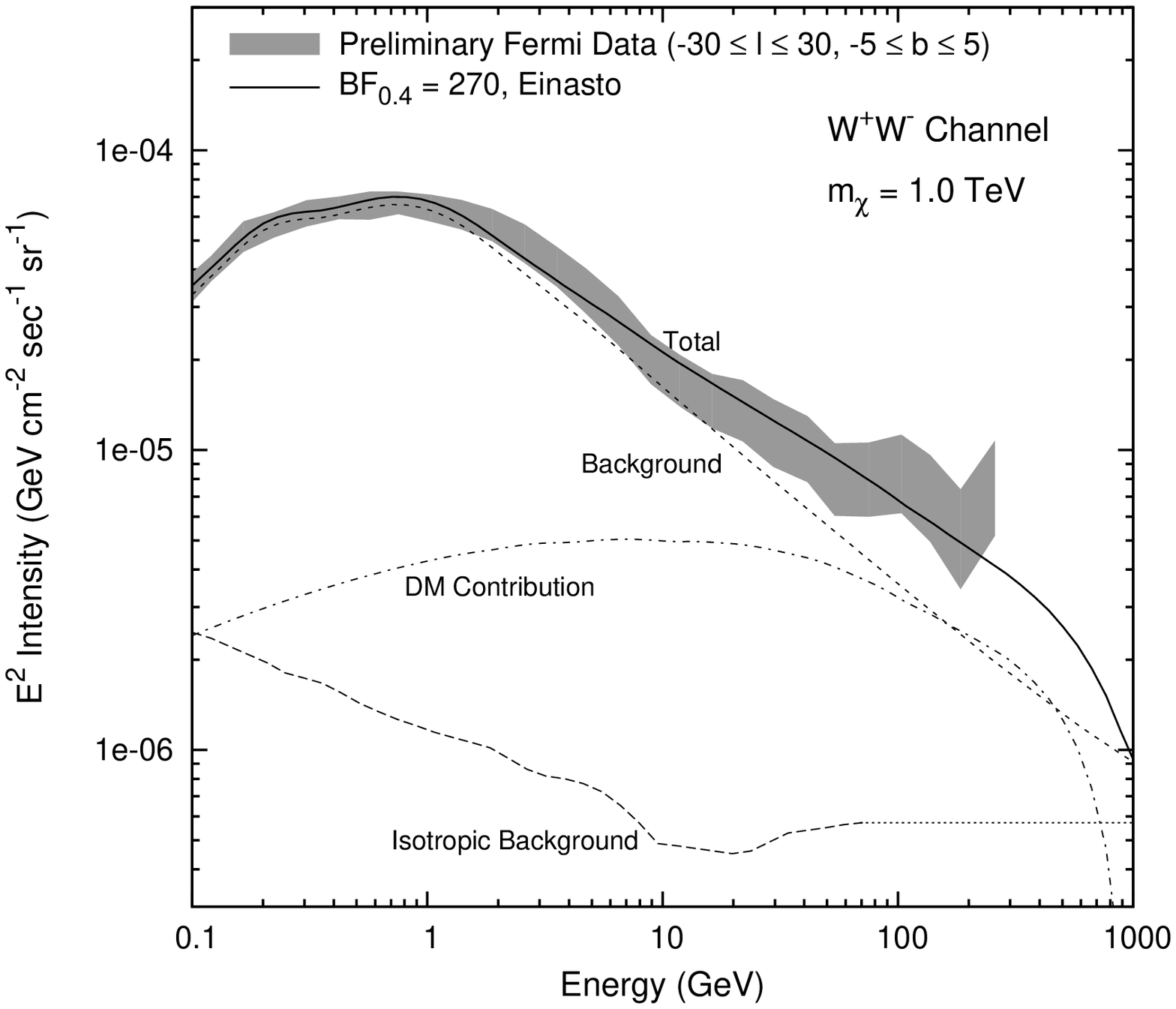}
\end{center}
\caption{The diffuse gamma ray signal in the inner galaxy  ($- 30^\circ < \ell < 30^\circ$, $-5^\circ < b <
  5^\circ$), with the DM signal arising principally from ICS in the top four.
\emph{Upper left}: Direct annihilation to muons, $\chi \chi \rightarrow \mu^+ \mu^-$.
$B_{0.4}$ is the boost factor required
relative to $\sigmav = 3\times10^{-26}\cmcubps$ and the reference
local DM density of $\rho_0 = 0.4 \GeV \cm^{-3}$.
\emph{Upper right}: XDM electrons, $\chi \chi \rightarrow \phi \phi$, followed by $\phi\rightarrow e^+e^-$.
\emph{Middle left}: XDM muons, $\chi \chi \rightarrow \phi \phi$, followed by $\phi\rightarrow \mu^+\mu^-$.
\emph{Middle right}:  XDM 1:1:2, $\chi \chi \rightarrow \phi \phi$, followed by $\phi\rightarrow \epp:\mu^+\mu^-:\pi^+\pi^-$ in a 1:1:2 ratio. The following two cases also have significant prompt photon contributions.
\emph{Lower left}: XDM taus, $\chi \chi \rightarrow \phi \phi$, followed by $\phi\rightarrow \tau^+\tau^-$.
\emph{Lower right}:  Direct annihilation to W's, $\chi \chi \rightarrow W^+W^-$.}
\label{fig:ICS}
\end{figure*}

Before proceeding to the plots of the IG gammas, we should again note caveats, both on the signal as well as data side. The predictions for the ICS photons depend dramatically on a number of inputs: propagation parameters, the ISRF, the magnetic field and the halo profile. The halo profile alone can change the signal by a significant amount, but even normalizing to the haze, there is still up to a factor of roughly 6 uncertainty in the ICS signal amplitude (see \cite{Cholis:2008wq,dobler:2009zz} for more details). Prompt photons suffer the usual halo uncertainty, but not the astrophysical ones. 

The data, it must be remembered, are preliminary. The highest energy data ($\gsim 100 \gev$) contain significant contamination from (non-photon) cosmic rays \cite{simonaprivate}, as we have discussed, and should the isotropic background sources rise, instead of staying flat, it could become important in this high energy range as well. Thus, rather than showing individual data points, we show a grey band which is an envelope of the preliminary allowed range. It should be expected to shift downward at the highest energies as the data are refined. Since our principal focus at this point is  how much signal could be seen in the inner region in light of these data, we shall not worry about the additional sources or contamination until they are fully studied.

We show in Figure \ref{fig:ICS} the signal for the inner galaxy region for which preliminary data have been shown recently by Fermi \cite{TROYTALK,DRELLTALK}. In viewing these plots, we see that ICS contributions do an excellent job of fitting the excess gamma rays. The highest energy data point is challenging to fit with ICS alone, however. We also show the XDM $\tau$ annihilation mode, which has prompt photons from $\pi^0$'s. The overall normalization of this channel is low compared to what would fit the local electronic excess, but demonstrates what even a small contribution of prompt hard photons can do to the spectrum. Such prompt photons could occur if $\phi$ has a subdominant decay to a $\pi^0$-rich mode, or if the DM annihilates into many $\phi$'s (as might occur in a non-Abelian model \cite{ArkaniHamed:2008qn,Baumgart:2009tn}. Likewise, if $m_\phi < 2 m_\mu$ and it is a scalar \cite{Cholis:2008vb,Nelson:2008hj}, then there can be an appreciable width into $\gamma \gamma$. 

\begin{figure*}
\begin{center}
\includegraphics[width=.45\textwidth]{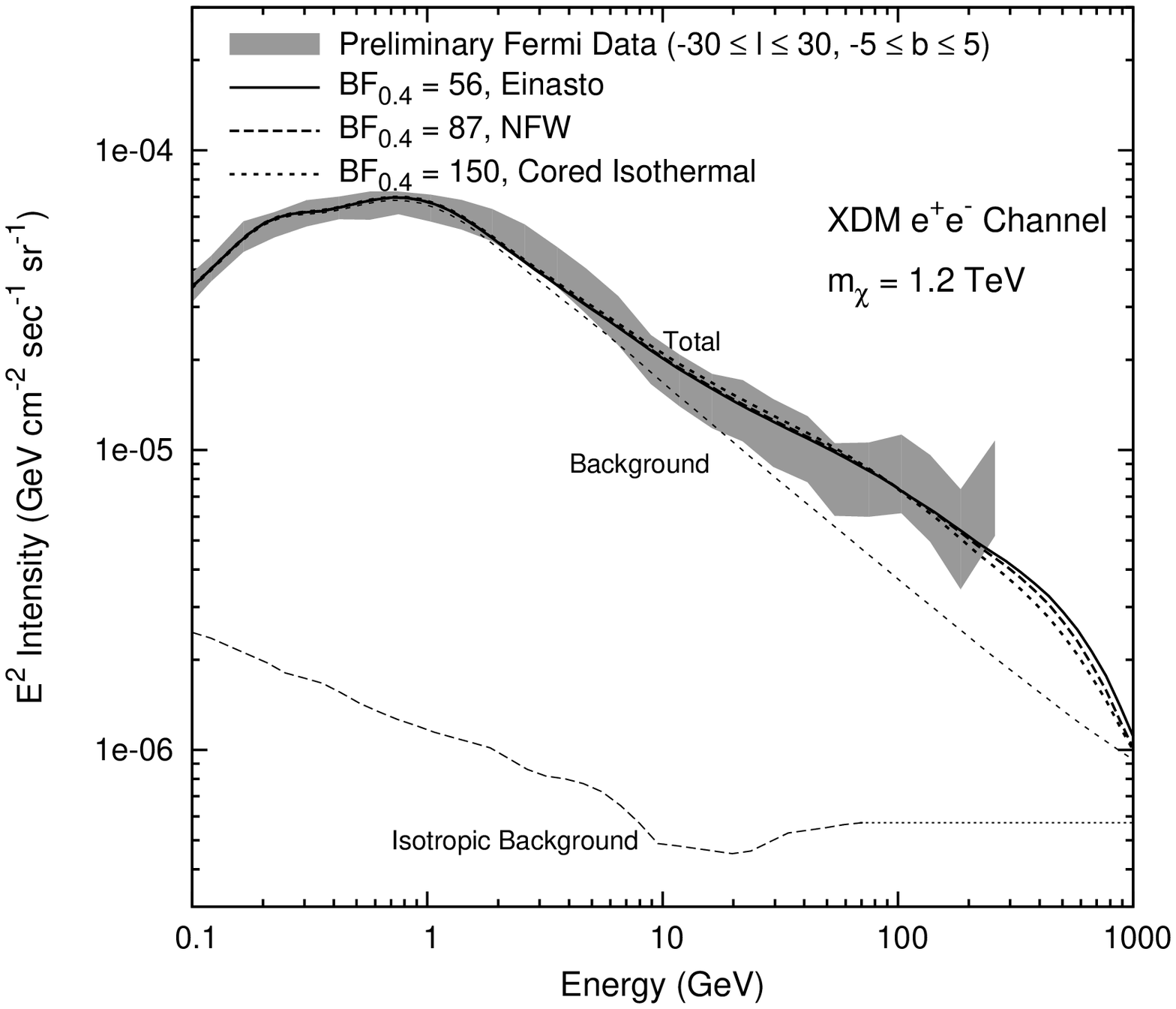}
\includegraphics[width=.45\textwidth]{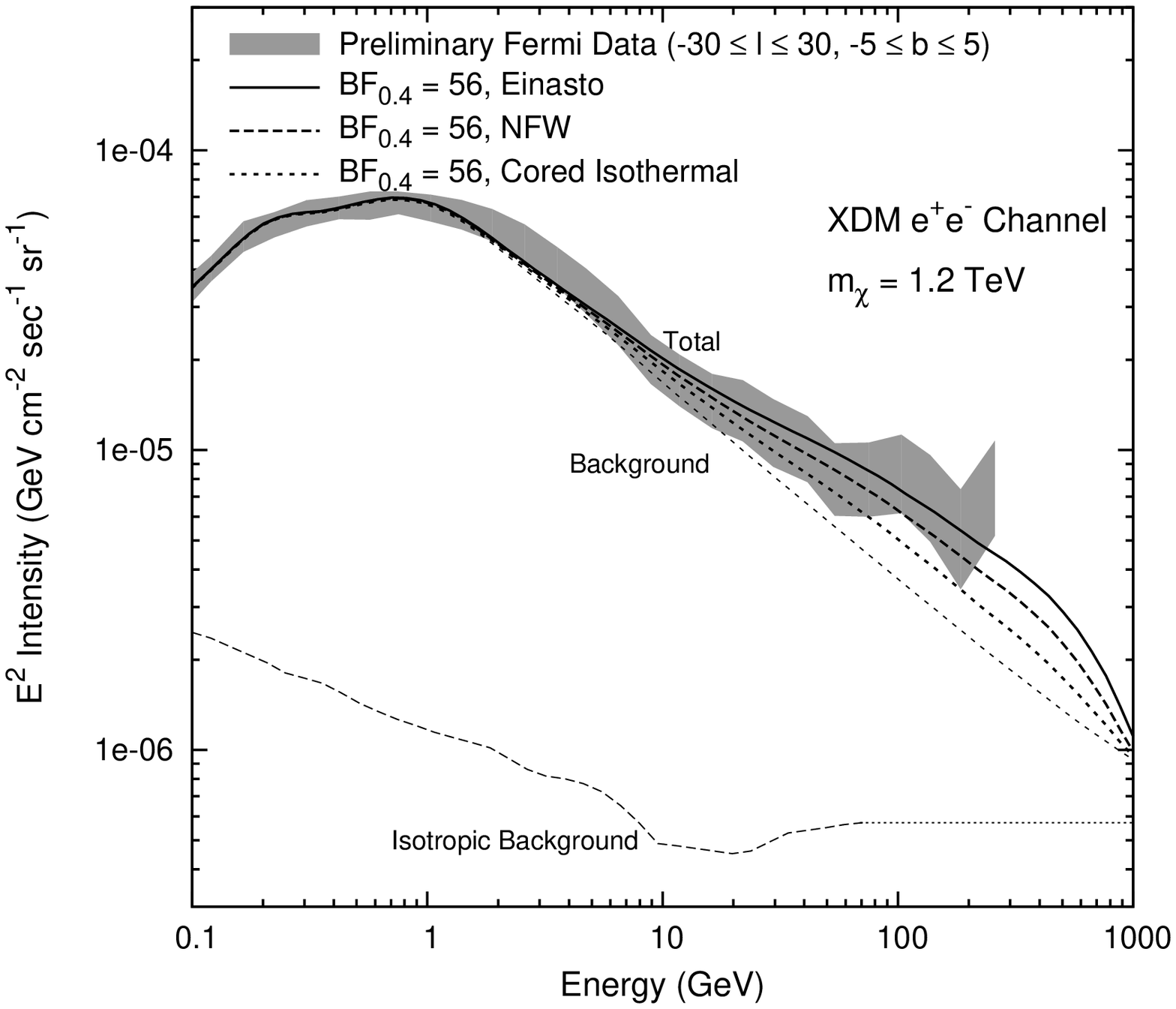}
\end{center}
\caption{The dependence of the inner Galaxy diffuse gamma ray signal on the dark matter density profile. {\em Left:} boost factors are chosen to give equal flux at 100 GeV, for the sample annihilation channel $\chi \chi \rightarrow \phi \phi$, followed by $\phi\rightarrow e^+e^-$. {\em Right:} Same mode, but constant boost factors.}
\label{fig:densityprofile}
\end{figure*}

The cross sections needed to fit the Fermi high energy electron excess (which we prefer to produce the normalization, as PAMELA is a ratio) \cite{Cholis:2008wq} are sometimes slightly larger than what is compatible with the Fermi IG data.  For the muon and XDM electron annihilation channels, the difference is roughly a factor of two, while for XDM muons and XDM 1:1:2 the boost factors are comparable \footnote{One must recall that the boosts used here are 0.56 smaller than in \cite{Cholis:2008wq} because of the difference in local density normalization  ($0.4~{\rm GeV/cm^3}$ vs. $0.3~{\rm GeV/cm^3}$).}. This, however, is well within the uncertainties, as the diffuse signal depends on, among other things, the halo profile (see figure \ref{fig:densityprofile}), the propagation parameters, and the contributors to the energy loss (specifically the magnetic field and ISRF). Thus, such close agreement is actually very encouraging. Such a difference may also arise from substructure (which may be disrupted in the galactic center) or from a change in the velocity dispersion (specifically, should it go up in the inner galaxy, as in \cite{RomanoDiaz:2008wz,RomanoDiaz:2009yq}).
\subsection{Varying the halo profile}
The only channels which seems to deviate noticeably from the preferred boost are the XDM tau channel, which is $\sim 6$ times lower than what is needed, and the $W^+W^-$ channel, which is roughly a factor of four lower. The  $W^+W^-$ channel has $\bar p$ limits which are already very significant \cite{Donato:2008jk}, and so likely would be excluded already. For the XDM tau channel, we must remember that the signal is dominated by the galactic center, and so a cored isothermal profile with a larger boost, for instance, should be able to give a similar signal with a boost comparable to what is needed to explain the Fermi/PAMELA $\epp$ signals. Such a soft core is required from HESS constraints in any event for this channel \cite{Bertone:2008xr,Bergstrom:2008ag,Meade:2009rb,Mardon:2009rc,Meade:2009iu}. Finally, as noted above, these channels, in particular the XDM tau mode, give some sense of what even a subdominant prompt photon contribution can do to the signal, and serve as a proxy for other modes. We see that the prompt contribution can be significant at high energies, even with a lower boost than the other channels. Should there be a prompt photon contribution, for instance from tau or kaon decay to $\pi^0$ (as might happen with a heavier $\phi$) it may be relevant, even if it is subdominant (for instance from kinematical suppression). 

Because the prompt and ICS photons have different distributions (see Figure \ref{fig:skydist}) a study of the distribution of the excesses as a function of energy should allow a distinction between prompt and ICS components, although this may be challenging. Additionally, such a prompt signal could be searched for in regions of the sky where the ICS signal is expected to be lower (for instance, the extragalactic emission or dwarf galaxies).

The density profile of dark matter in the inner galaxy is not well known. Cosmological numerical simulations of cold dark matter predict halo density profiles with cuspy inner regions, well described by the Einasto \cite{Merritt:2005xc} or Navarro-Frenk-White (NFW) \cite{Navarro:1995iw} profiles. However, the inclusion of baryons may flatten out these cusps, yielding halos better modeled by an isothermal sphere (see, e.g., \cite{RomanoDiaz:2008wz,RomanoDiaz:2009yq}). 

To test the effect of varying the dark matter density profile, we performed a sample calculation using three different halo models :
\begin{align} \rho(r) = & \rho_0 \frac{r_c}{r} \frac{1}{(1 + r/r_c)^2} & \mathrm{NFW},  \\
\rho(r) = & \rho_0 \exp\left(-\frac{2}{\alpha} \left(\frac{r^\alpha - R_\odot^\alpha}{r_{-2}^\alpha} \right) \right) & \mathrm{Einasto},  \\ \rho(r) = & \rho_0 \frac{r_c^2 + R_\odot^2}{r_c^2 + r^2} & \mathrm{Cored Isothermal} .\end{align}
Here $R_\odot = 8.5$ kpc is the solar distance from the Galactic center,
$r$ is the spherical radial coordinate, $r_c$ is the core radius and
$\rho_0$ is the local value of the dark matter mass density, except in
the NFW case, where $\rho(r_c) = \rho_0/4$. For the NFW profile we have taken $r_{c} = 20$ kpc. For the Einasto profile we have used $\alpha = 0.17$, with $r_{-2} = 25$ kpc, while the ``cored isothermal'' profile used here is taken from \cite{Moskalenko:1999sb}, with $r_{c} = 2.8$ kpc. 

Figure \ref{fig:densityprofile} shows the effect of varying the dark matter density profile on the gamma ray spectrum and the boost factor required to fit the data. The required boost factor is sensitive to the choice of profile, but the spectral shape is nearly unchanged. The largest boost factors are required for the isothermal profile due to its lower density in the inner galaxy; Einasto and NFW halos give rise to more annihilation and hence require smaller boosts.
\subsection{Predictions for the Galactic Center and ``Four Corners''}
\begin{figure*}
\begin{center}
\includegraphics[width=.45\textwidth]{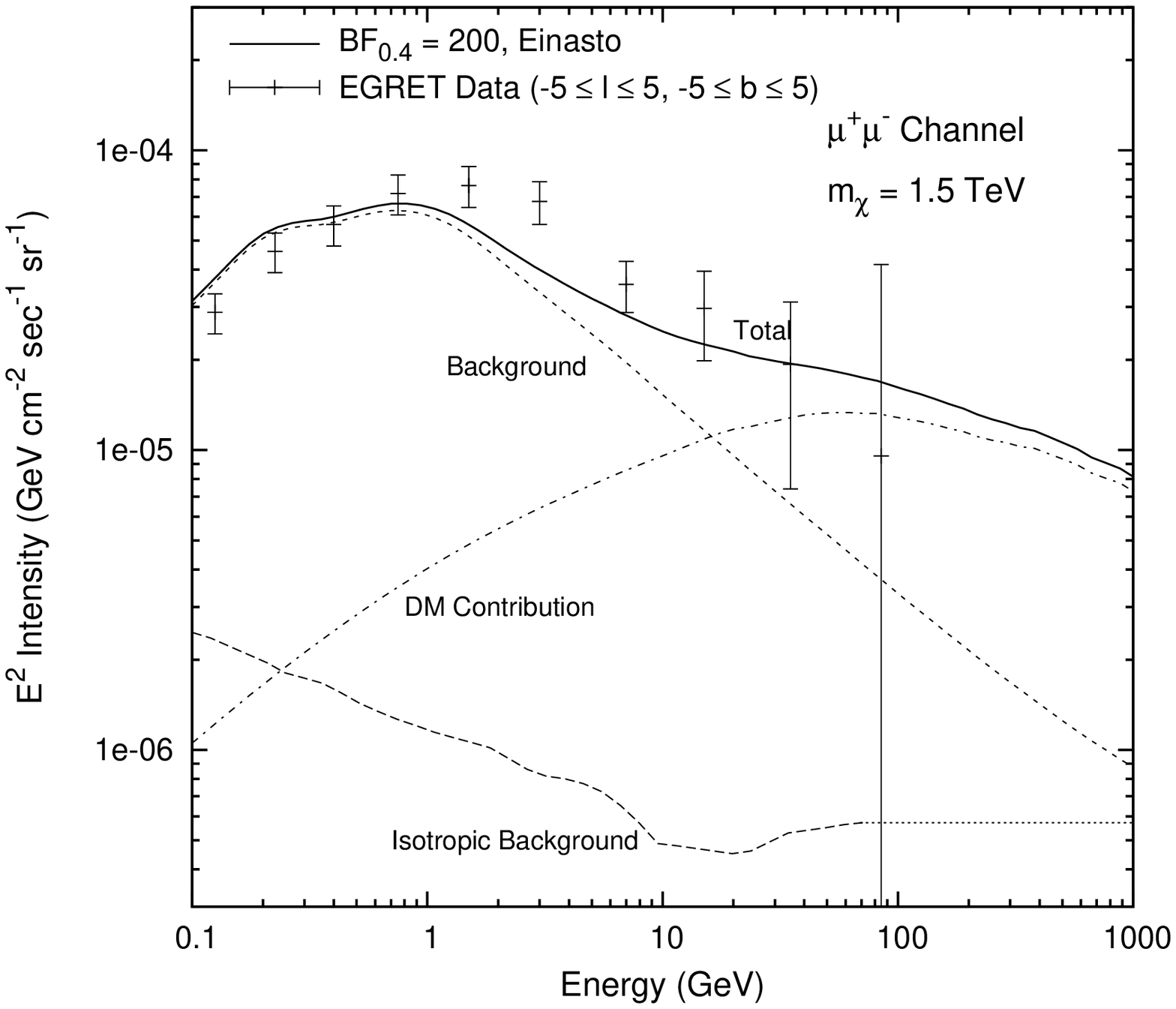}\hskip 0.2in
\includegraphics[width=.45\textwidth]{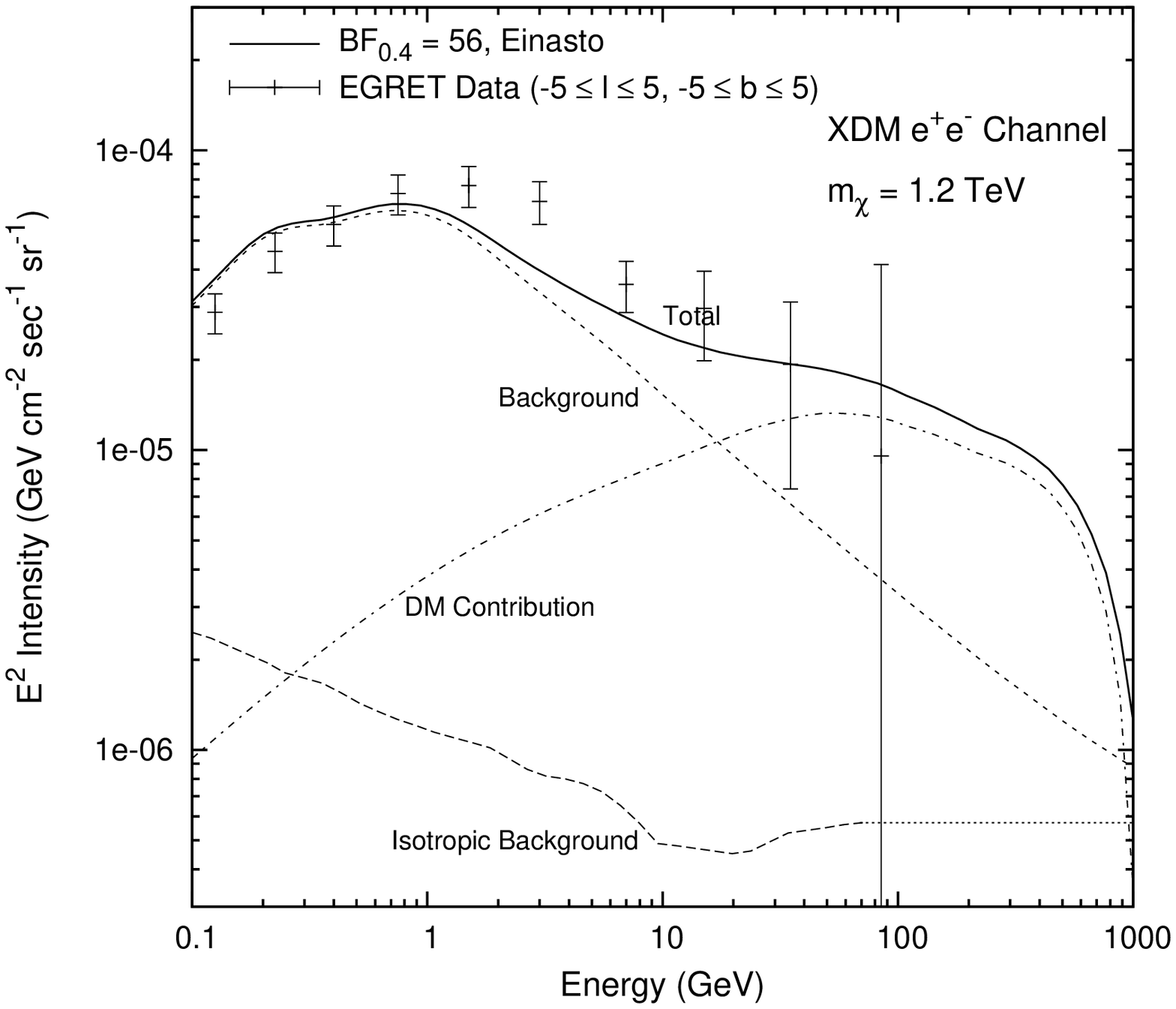}\\
\includegraphics[width=.45\textwidth]{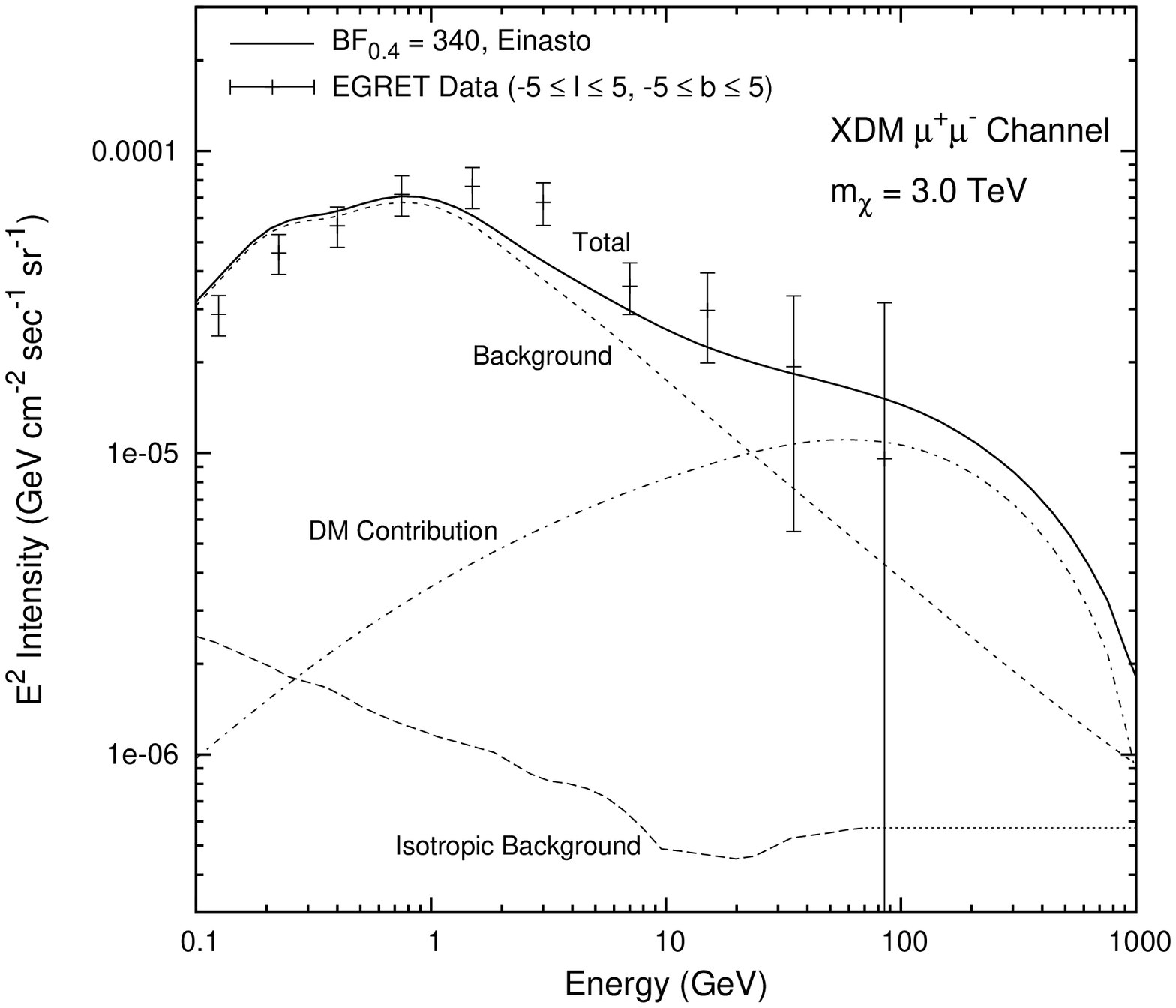}\hskip 0.2in
\includegraphics[width=.45\textwidth]{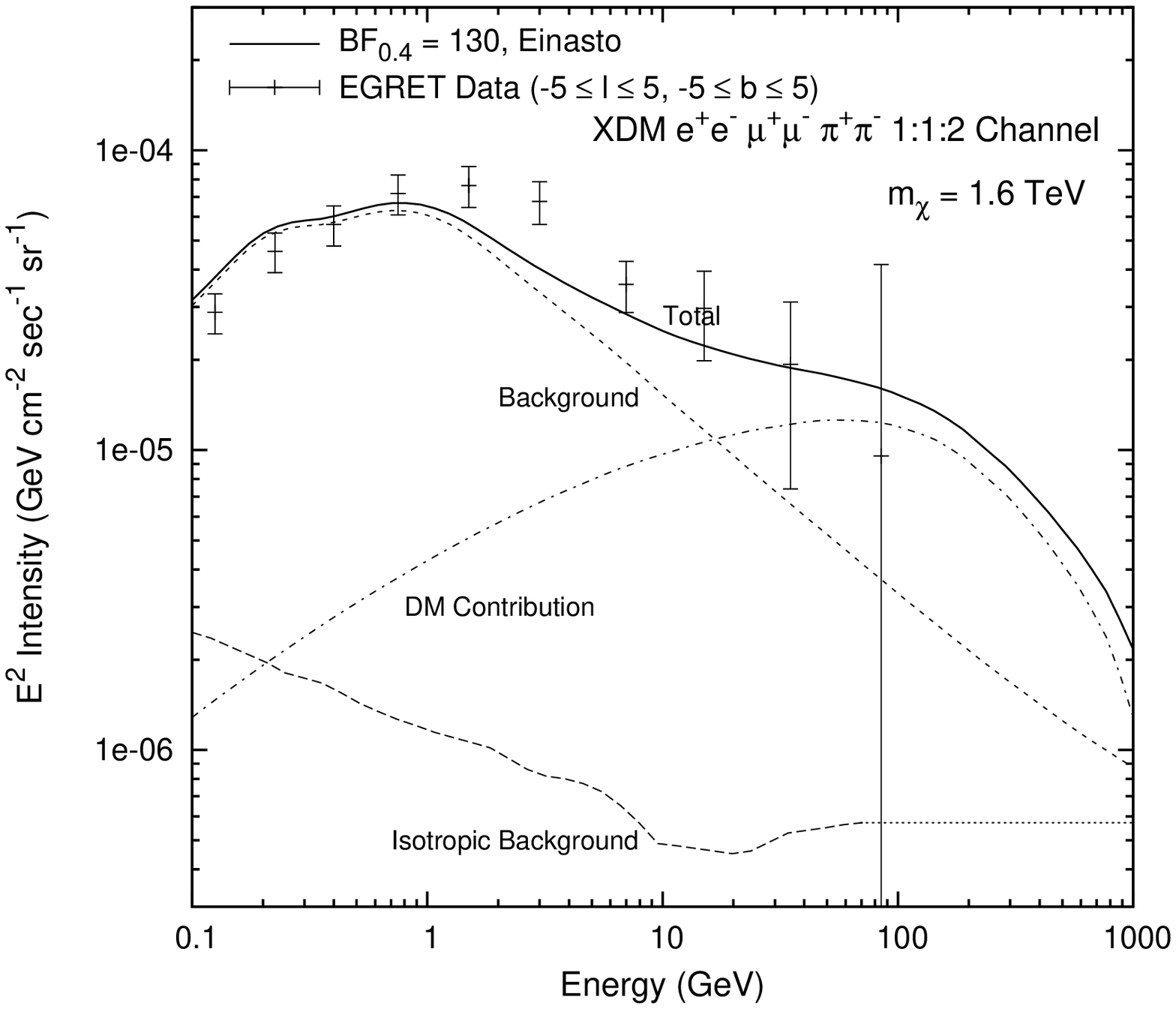}
\includegraphics[width=.45\textwidth]{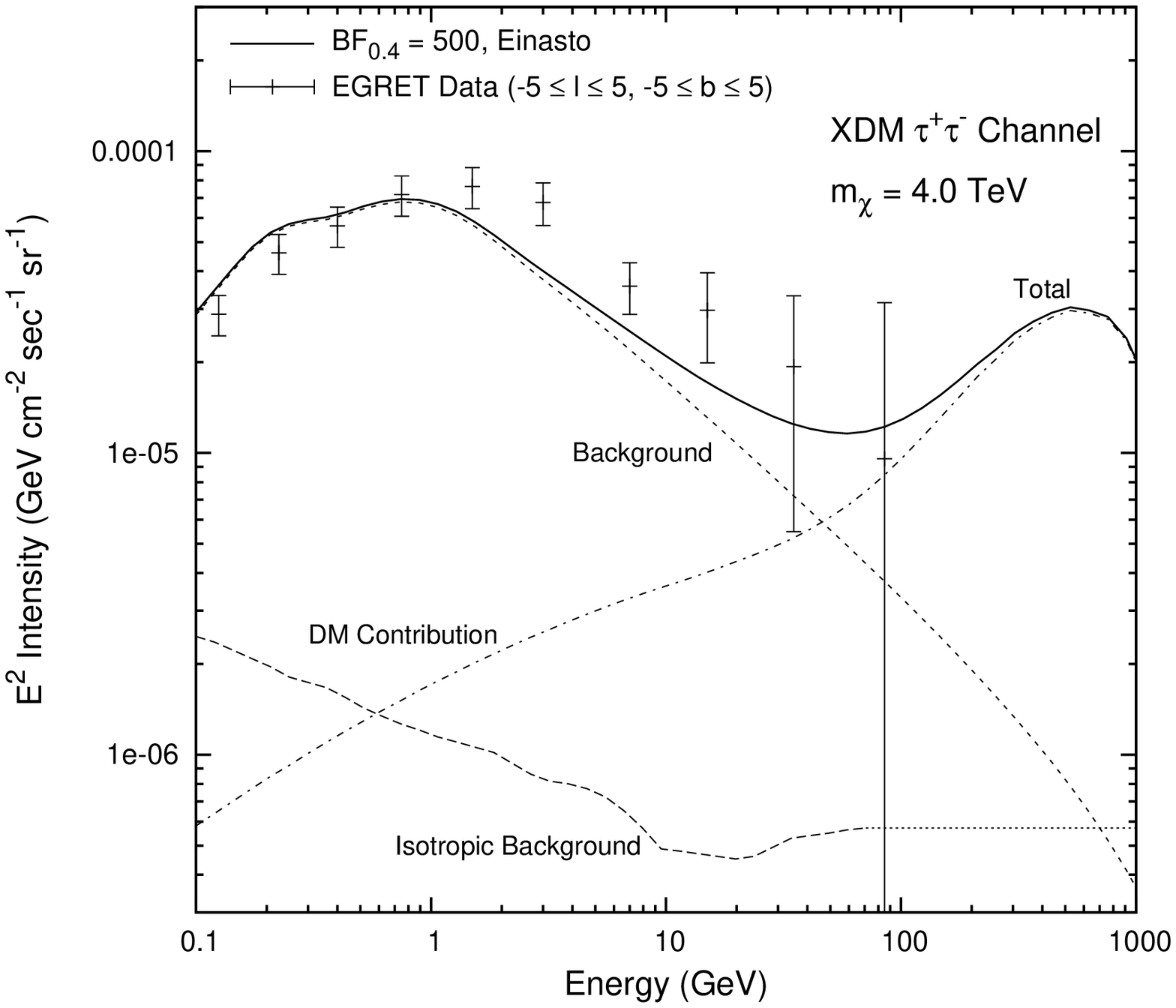}\hskip 0.2in
\includegraphics[width=.45\textwidth]{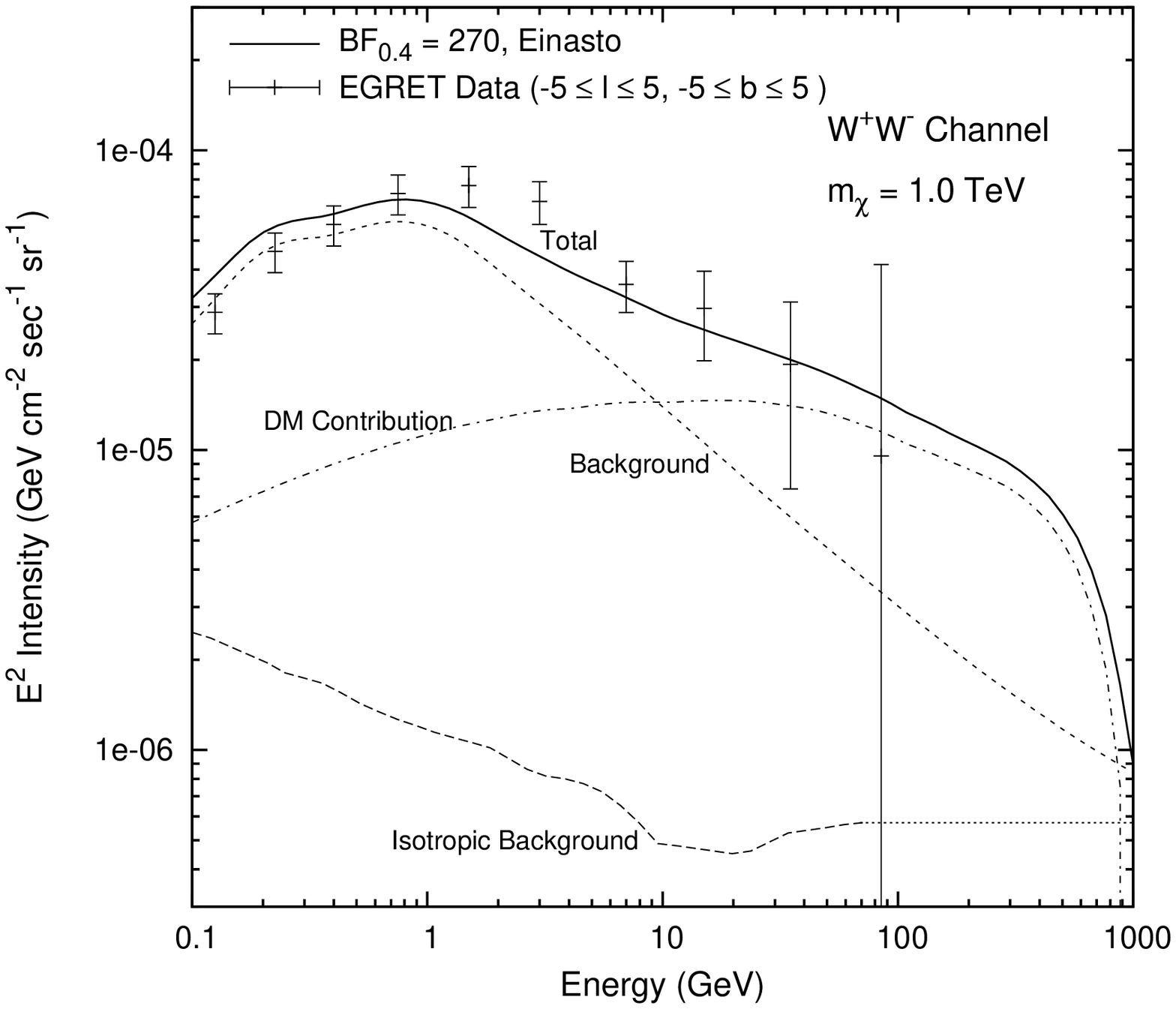}
\end{center}
\caption{The gamma ray signal in the galactic center ($- 5^\circ < \ell < 5^\circ$, $-5^\circ < b <  5^\circ$) normalized to the IG.
\emph{Upper left}: Direct annihilation to muons, $\chi \chi \rightarrow \mu^+ \mu^-$.
$B_{0.4}$ is the boost factor required
relative to $\sigmav = 3\times10^{-26}\cmcubps$ and the reference
local DM density of $\rho_0 = 0.4 \GeV \cm^{-3}$.
\emph{Upper right}: XDM electrons, $\chi \chi \rightarrow \phi \phi$, followed by $\phi\rightarrow e^+e^-$.
\emph{Middle left}: XDM muons, $\chi \chi \rightarrow \phi \phi$, followed by $\phi\rightarrow \mu^+\mu^-$.
\emph{Middle right}:  XDM 1:1:2, $\chi \chi \rightarrow \phi \phi$, followed by $\phi\rightarrow \epp:\mu^+\mu^-:\pi^+\pi^-$ in a 1:1:2 ratio.
\emph{Lower left}: XDM taus, $\chi \chi \rightarrow \phi \phi$, followed by $\phi\rightarrow \tau^+\tau^-$.
\emph{Lower right}:  Direct annihilation to W's, $\chi \chi \rightarrow
W^+W^-$.
Data points are from the Strong et
al. re-analysis of the EGRET data \cite{Strong:2005zx}, which found a
harder spectrum at $10-100$ GeV within a few degrees of the GC, using
improved sensitivity estimates from \cite{Thompson:2005}.
}
\label{fig:inner5}
\end{figure*}

Figure \ref{fig:skydist} shows the contribution to the gamma ray spectrum as a function of angle from the GC. We see that most of the contribution is from the center, and that signal falls off more rapidly than background. Hence, moving to the galactic center should provide a stronger signal compared to background, so long as one is away from the true GC, where significant point sources can dominate. We show in Figure \ref{fig:inner5} the expected signal for the models for the inner $5^\circ$ of the galaxy. We see that based upon the signal which could possibly be in the IG, a significant signal should also be seen in this region. Because of the aforementioned uncertainties, the signal could be significantly lower, by possibly a factor of six, even normalizing to the haze \cite{Cholis:2008wq,dobler:2009zz}.

The amplitude of the signal is also uncertain, however, because most of the signal arises from the inner $10^\circ$ for an Einasto profile, and in fact, mostly from the inner $5^\circ$. Thus uncertainties in the halo profile clearly will affect the size of the GC-$5^\circ$ signal as well. At the same time, even for the cored isothermal halo, any significant contribution to the IG is coming from the inner region, and so normalizing to a contribution allowed in the present data, a significant signal could be present.

In the innermost $1^\circ$ of the Galaxy, initial measurements from the Fermi LAT can place constraints on the possible dark matter annihilation cross section \cite{MEURERTALK}. The total integrated gamma ray flux, including background, between 100 MeV and 100 GeV is measured to be $1.22 \times 10^{-6}$ cm$^{-2}$ s$^{-1}$ (no attempt has been made to subtract background at this stage). The signals from our models in this region, with the boost factors required to fit the IG gamma ray spectrum, are uniformly well below this limit -- as would be expected, since the Galactic center is expected to be highly background dominated. For example, for the case of a 2.5 TeV WIMP annihilating through the ``XDM muons'' channel ($\chi \chi \rightarrow \phi \phi$ followed by $\phi \rightarrow \mu^+ \mu^-$), the integral flux over 100 MeV -- 2 TeV is $\sim 3 (\mathrm{BF} / 1000) \times 10^{-7}$ cm$^{-2}$ s$^{-1}$ for an Einasto halo profile.

\begin{figure*}
\begin{center}
\includegraphics[width=.45\textwidth]{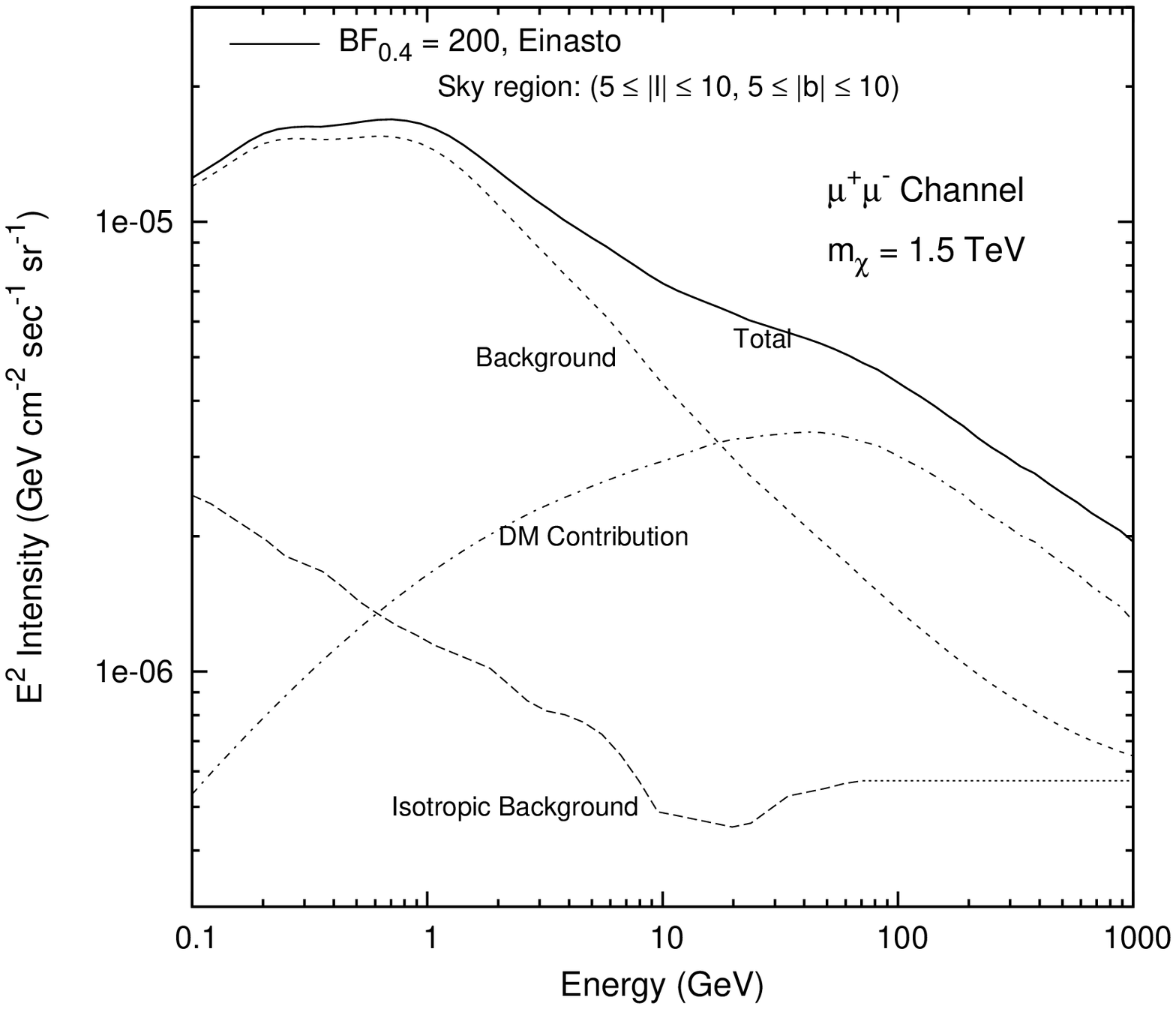}\hskip 0.2in
\includegraphics[width=.45\textwidth]{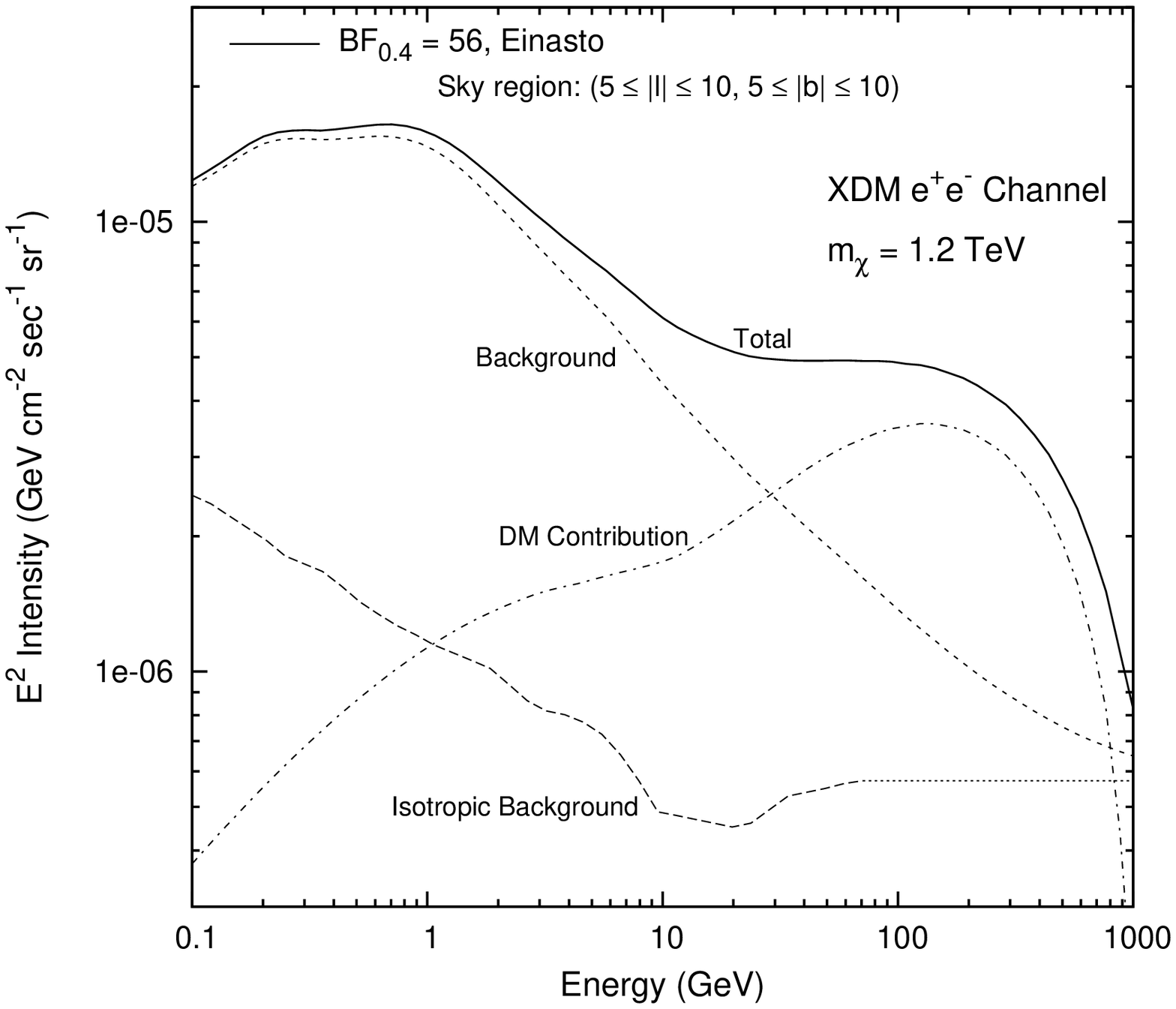}\\
\includegraphics[width=.45\textwidth]{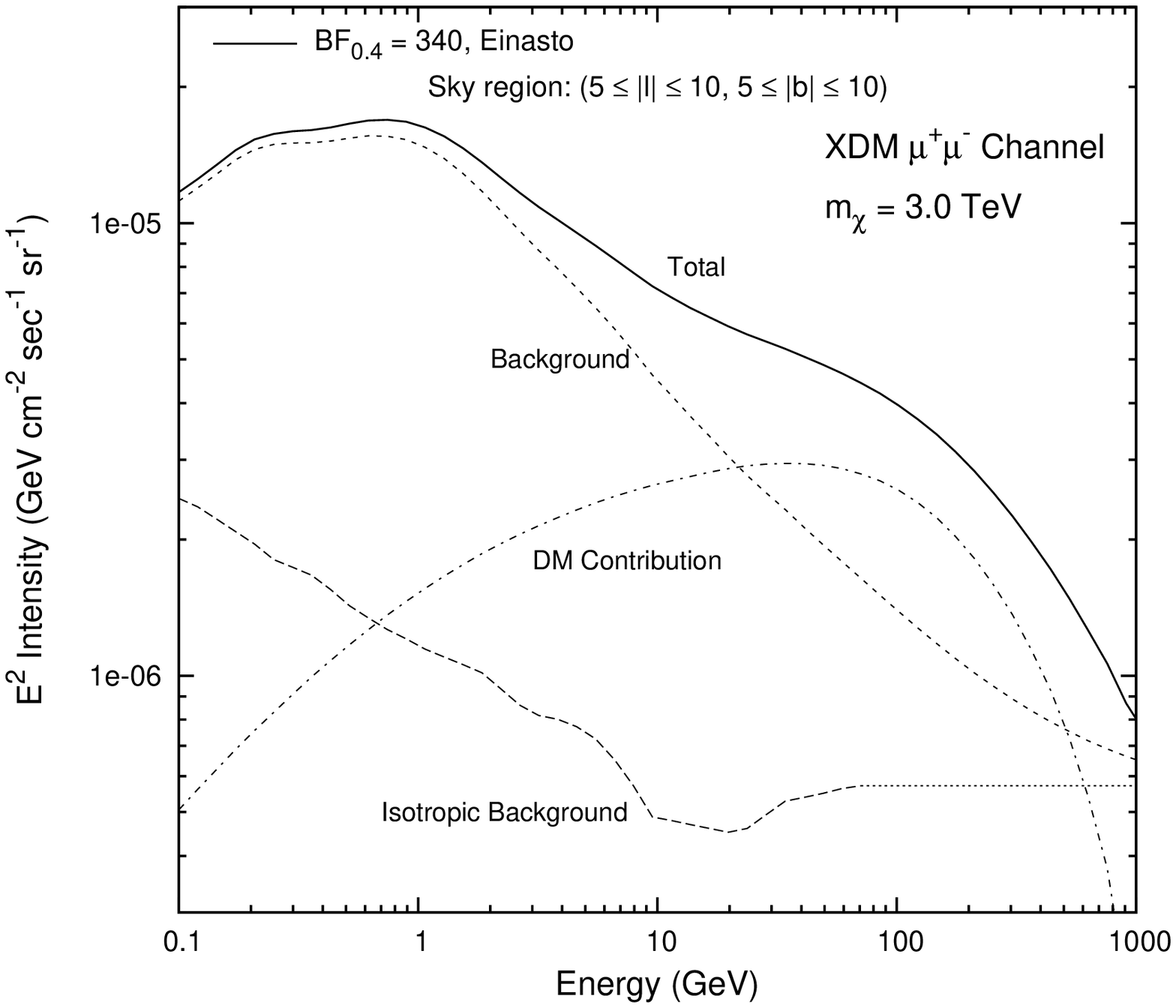}\hskip 0.2in
\includegraphics[width=.45\textwidth]{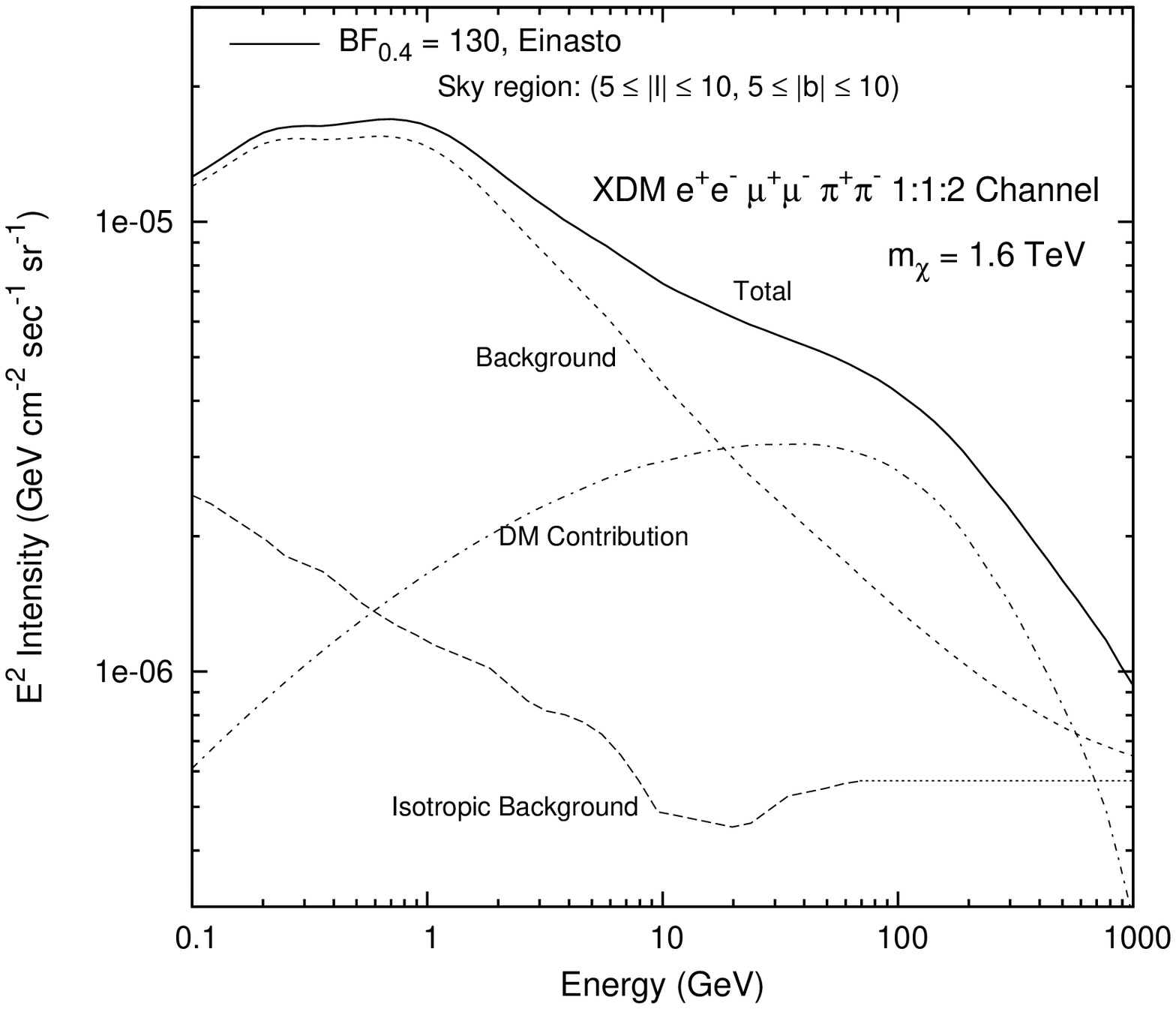}
\includegraphics[width=.45\textwidth]{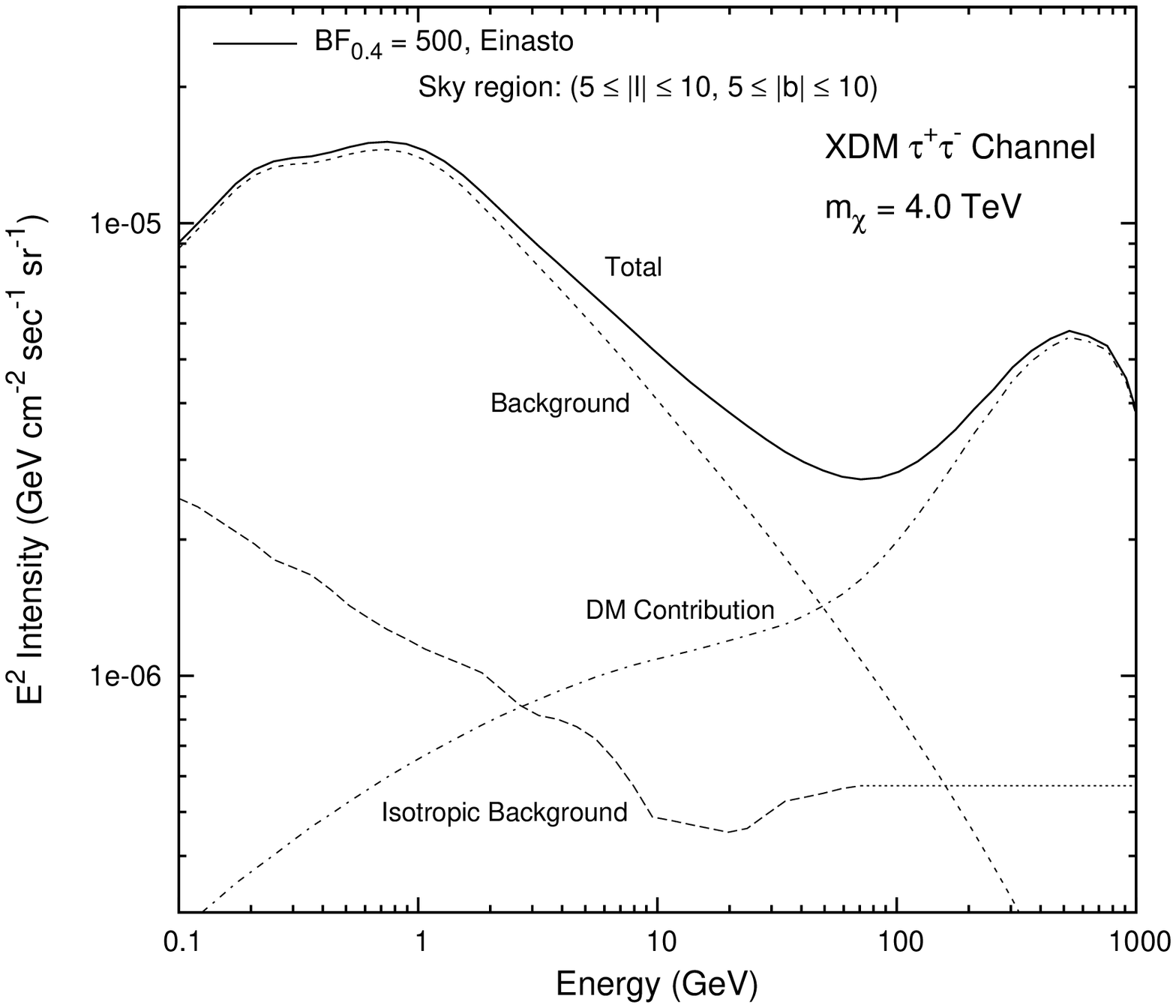}\hskip 0.2in
\includegraphics[width=.45\textwidth]{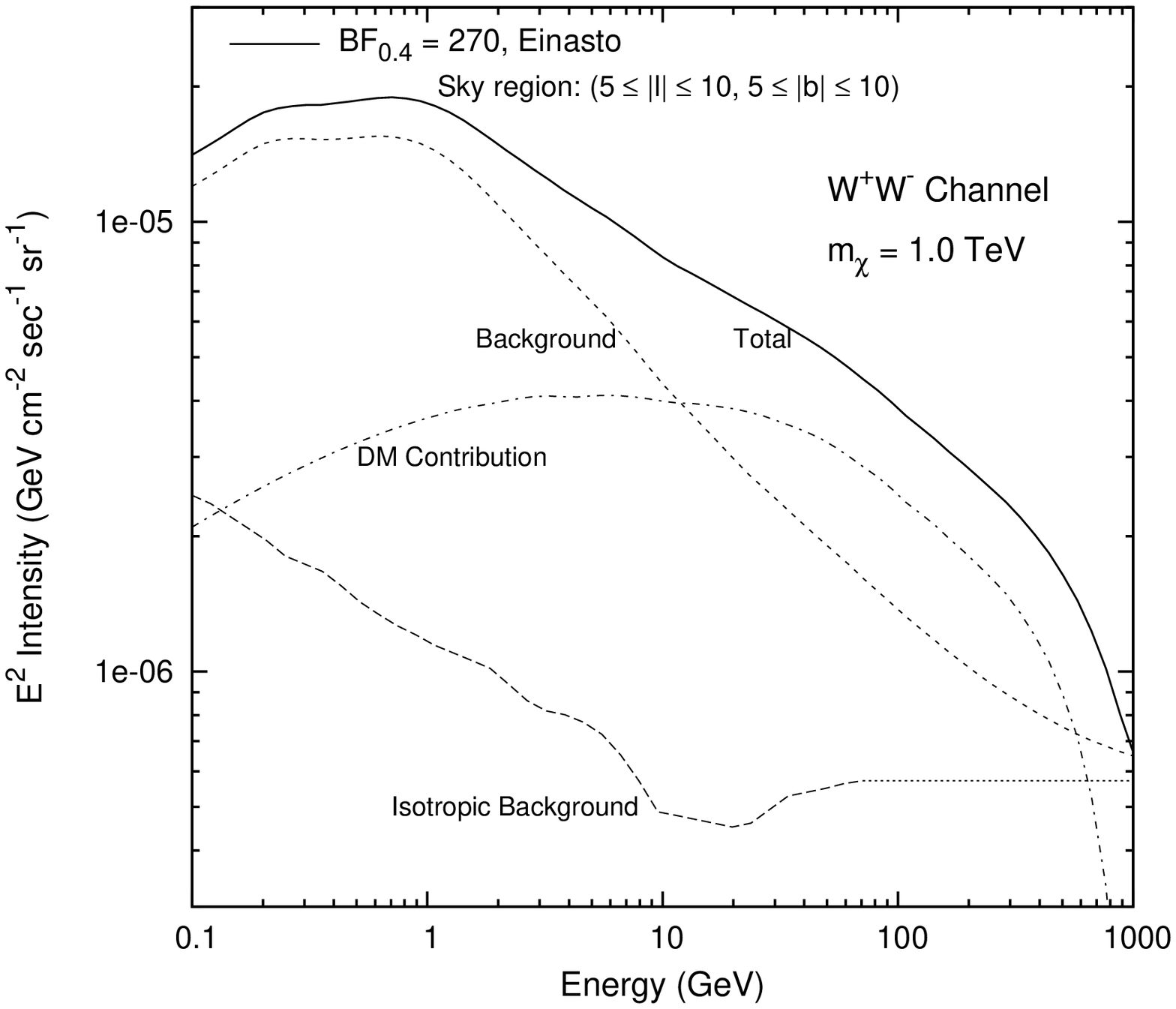}
\end{center}
\caption{The gamma ray signal in the four corners region  ($5^\circ < |\ell| < 10^\circ$, $5^\circ < |b| <  10^\circ$), normalized to the IG.
\emph{Upper left}: Direct annihilation to muons, $\chi \chi \rightarrow \mu^+ \mu^-.$
$B_{0.4}$ is the boost factor required
relative to $\sigmav = 3\times10^{-26}\cmcubps$ and the reference
local DM density of $\rho_0 = 0.4 \GeV \cm^{-3}$.
\emph{Upper right}: XDM electrons, $\chi \chi \rightarrow \phi \phi$, followed by $\phi\rightarrow e^+e^-$.
\emph{Middle left}: XDM muons, $\chi \chi \rightarrow \phi \phi$, followed by $\phi\rightarrow \mu^+\mu^-$.
\emph{Middle right}:  XDM 1:1:2, $\chi \chi \rightarrow \phi \phi$, followed by $\phi\rightarrow \epp:\mu^+\mu^-:\pi^+\pi^-$ in a 1:1:2 ratio.
\emph{Lower left}: XDM taus, $\chi \chi \rightarrow \phi \phi$, followed by $\phi\rightarrow \tau^+\tau^-$.
\emph{Lower right}:  Direct annihilation to W's, $\chi \chi \rightarrow W^+W^-$.}
\label{fig:4C}
\end{figure*}

Of course, the galactic center is a challenging place to look for
signal. There are numerous point sources, and the disk of the galaxy is
very bright.  Near $\ell = 0^\circ$, it is difficult to map galactic
rotation to distance, so structures cannot be placed along the line of
sight. Such problems can be alleviated by moving off the plane by
$5^\circ$ in $b$ and away from the center by $5^\circ$ in $\ell$. This
motivates a consideration of the ``four corners'' (FC) region, defined
to be ($5^\circ < |\ell| < 10^\circ$, $5^\circ < |b| < 10^\circ$). Such
a region has lower backgrounds and, additionally, lower uncertainties in
its background. In some sense this is taking the idea of an annulus,
proposed previously \cite{Hooper:2007gi} a step further. We show in
Figure \ref{fig:4C} the signal expected in this region of the sky. While
the total signal is smaller than in the inner $5^\circ$, the lower
uncertainties and high S/B there should make the signal more pronounced.

As one can see the S/B is {\em very} high in this region. Although there are still uncertainties in e.g., diffusion parameters and the magnetic field, the strength of this signal should give conclusive evidence for these models.
\section{Discussion}
As Fermi moves into the central regions of the galaxy, the possibility of detection of a dark matter signal increases. Already the present IG data places limits on how strong a signal could be, but, interestingly, is still compatible with a contribution expected from ICS signals of dark matter models which explain local $\epp$ excesses. As the data are refined, a broad excess could be indicative of an ICS signal, while a peakier signal at higher energies is more easily associated with a prompt photon contribution.

As these data are preliminary, and the region of the sky is dominated by background, one must be extremely cautious, and constraints will likely strengthen. Still, as the DM interpretation of the electronic excess has such a robust ICS prediction, it is certainly exciting that such a signal could be present at the expected size. As the majority of the DM signal in the IG originates dominantly from the inner part of the region considered, as Fermi narrows the angular size, the significance should increase, as long as point sources in the GC are treated appropriately. The most convincing data may come from a four-corners analysis, where backgrounds are better understood. In any event, in the near future, data releases from Fermi should shed significant light on the nature of dark matter.

\vskip 0.2 in
\noindent {\bf Acknowledgments}
\vskip 0.05in
\noindent We thank Nima Arkani-Hamed, Igor Moskalenko, Simona Murgia and Michele Papucci for useful conversations. This work was partially supported by the Director, Office of Science,
of the U.S.  Department of Energy under Contract
No. DE-AC02-05CH11231. NW is supported by NSF CAREER grant PHY-0449818, and IC, LG and NW are supported by DOE OJI grant \# DE-FG02-06ER41417.



\bibliography{ics}
\bibliographystyle{apsrev}

\end{document}